\documentclass[tunneling,twocolumn,showpacs,superscriptaddress,floatfix]{revtex4}
\usepackage{graphicx}
\usepackage{psfrag}
\usepackage{color}

\usepackage{amsmath} 
\usepackage{amssymb}   
\usepackage{amstext}
\usepackage{amsthm}
\usepackage{amsfonts}
\usepackage{amsxtra}

\begin{document}

\newcommand{\ket}[1]{\left\vert #1\,\right\rangle}
\newcommand{\ketrc}{\left\vert\textsc{rc}\,\right\rangle}
\newcommand{\bra}[1]{\left\langle #1\,\right\vert}
\newcommand{\bbra}[1]{\left\langle\left\langle #1\,\right\vert\right.}
\newcommand{\brarc}{\left\langle\,\textsc{rc}\right\vert}
\newcommand{\braket}[2]{\left\langle #1\,\vert #2\,\right\rangle}
\newcommand{\bbraket}[2]{\left\langle\left\langle #1\,\vert #2\,\right\rangle\right.}
\newcommand{\de}{\partial}
\newcommand{\eps}{\varepsilon}
\newcommand{\tr}{\operatorname{\mathrm{Tr}}}
\newcommand{\re}{\mathrm{Re}}
\newcommand{\im}{\mathrm{Im}}
\newcommand{\R}{\mathbb{R}}
\newcommand{\eav}[1]{\left\langle #1\right\rangle}
\newcommand{\beq}{\begin{equation}}
\newcommand{\eeq}{\end{equation}}
\newcommand{\ben}{\begin{eqnarray}}
\newcommand{\een}{\end{eqnarray}}
\newcommand{\bea}{\begin{array}}
\newcommand{\eea}{\end{array}}
\newcommand{\om}{(\omega )}
\newcommand{\bef}{\begin{figure}}
\newcommand{\eef}{\end{figure}}
\newcommand{\leg}[1]{\caption{\protect\rm{\protect\footnotesize{#1}}}}
\newcommand{\ew}[1]{\langle{#1}\rangle}
\newcommand{\be}[1]{\mid\!{#1}\!\mid}
\newcommand{\no}{\nonumber}
\newcommand{\etal}{{\em et~al }}
\newcommand{\geff}{g_{\mbox{\it{\scriptsize{eff}}}}}
\newcommand{\da}[1]{{#1}^\dagger}
\newcommand{\cf}{{\it cf.\/}\ }
\newcommand{\ie}{{\it i.e.\/}\ }   
\setlength\abovedisplayskip{5pt}
\setlength\belowdisplayskip{5pt}

\title{Shielding and localization 
  in presence of  long range hopping}
\author{G.~L.~Celardo}
\affiliation{Dipartimento di Matematica e
Fisica and Interdisciplinary Laboratories for Advanced Materials Physics,
 Universit\`a Cattolica, via Musei 41, 25121 Brescia, Italy,\\
 Istituto Nazionale di Fisica Nucleare,  Sezione di Pavia, 
via Bassi 6, I-27100,  Pavia, Italy}
\affiliation{Center for Theoretical Physics of Complex Systems, Institute for Basic Science, Daejeon, Korea}
\affiliation{Instituto de F\'isica, Benem\'erita Universidad Aut\'onoma de Puebla, Apartado Postal J-48, Puebla 72570, Mexico}
\author{R.~Kaiser}
\affiliation{Universit\'e de Nice Sophia Antipolis, CNRS, Institut
  Non-Lin\'eaire de Nice, UMR 7335, Valbonne F-06560, France}

\author{F.~Borgonovi}
\affiliation{Dipartimento di Matematica e
Fisica and Interdisciplinary Laboratories for Advanced Materials Physics,
 Universit\`a Cattolica, via Musei 41, 25121 Brescia, Italy
and\\
 Istituto Nazionale di Fisica Nucleare,  Sezione di Pavia, 
via Bassi 6, I-27100,  Pavia, Italy}
                                                        
\begin{abstract}    
We investigate a paradigmatic model for quantum transport with both
nearest-neighbor and infinite range hopping coupling (independent of
the position). Due to
long range homogeneous hopping, a gap between the ground state and the excited
states  can be induced, which is  mathematically equivalent to the
superconducting gap. In the gapped regime, the dynamics within the
excited states subspace  is shielded from long range hopping, namely
it occurs as if long range hopping would be absent.
This is a cooperative phenomenon  since shielding is effective  over a 
time scale which diverges with the system size.  
We named this effect {\it Cooperative Shielding}. 
We  also discuss the consequences of our findings on Anderson
localization. Long range hopping is usually thought to destroy 
localization due to the fact that it induces an infinite number of
resonances. Contrary to this common lore
we show that the excited states display strong localized
features when shielding is effective even in the regime of
strong long range coupling.  A brief discussion on the
extension of our results to generic power-law decaying long range
hopping is also given. Our preliminary results confirms that the
effects found for the infinite range case are generic.
%On the other hand, the ground state shows
%robustness to disorder since it localizes above a critical disorder strength
%proportional to the system size. 
\end{abstract}                                                               
                                                                                                         
\date{\today}                               
\maketitle

\maketitle

\section{Introduction}

In recent years, technological advancement allowed to engineer
several systems in which the role of quantum coherence is
essential to understand their dynamics.  
In view of these considerations, searching for  novel coherent effects 
%which are robust to disorder 
is  fundamental to exploit quantum
properties in  technological devices such as quantum wires, quantum
computers, quantum sensors.
Of particular interest is the topic of transport of energy or charge
in the quantum coherent regime, due to its relevance in many technological
applications, such as in light-harvesting systems~\cite{fmo}, molecular wires~\cite{exciton}
and in other mesoscopic systems~\cite{meso}.

Recently, great attention has been devoted to quantum transport in models with long range
interactions due to their  relevance in many condensed matter physical systems. 
Indeed long range interactions between the constituents of a system do not
arise only from microscopic interactions, but in many 
condensed matter systems they can be
induced by the coupling with environmental modes having a
wavelength larger than  the system size. This "mediated" long range
interaction arises in several systems:
 in ion traps~\cite{iontrap} due
to the coupling of the trapped ions  with large
wave-length phonon modes;
in cold atoms~\cite{kaiser} and in natural light harvesting
systems~\cite{fmo}, due to the coupling
with the electromagnetic field (EMF) when the wave length of
the photon is much larger than the system size.
Long range interacting systems display particular features that are
not often observed in other systems, such as broken ergodicity~\cite{tnt}, and long-lasting out-of-equilibrium regimes~\cite{romain}.
%A major topic of interest has been whether the propagation of
%excitations in systems with long range interaction remains or not
%confined to an effective light cone~\cite{lightcone}, as defined by
%the Lieb-Robinson bound~\cite{lieb} and its generalisations (
%Ref.~\cite{lieb2} and references therein).
Out of equilibrium dynamics of such models have been widely analyzed
both experimentally~\cite{iontrap} and
theoretically~\cite{kastner,lea}, showing non trivial
cooperative effects and strong dependence of the dynamical evolution
on the initial state. Together with a very fast  spreading of
information~\cite{iontrap} and the destruction of localization~\cite{levitov}, also the opposite behaviour has been reported in case of long range
interacting system: the suppression of  information
spreading~\cite{kastner,selftrapping} and strong signatures of
localization~\cite{ossipov,alberto,Giulio}.

\begin{figure}[t!] 
\centering
\includegraphics[width=\columnwidth]{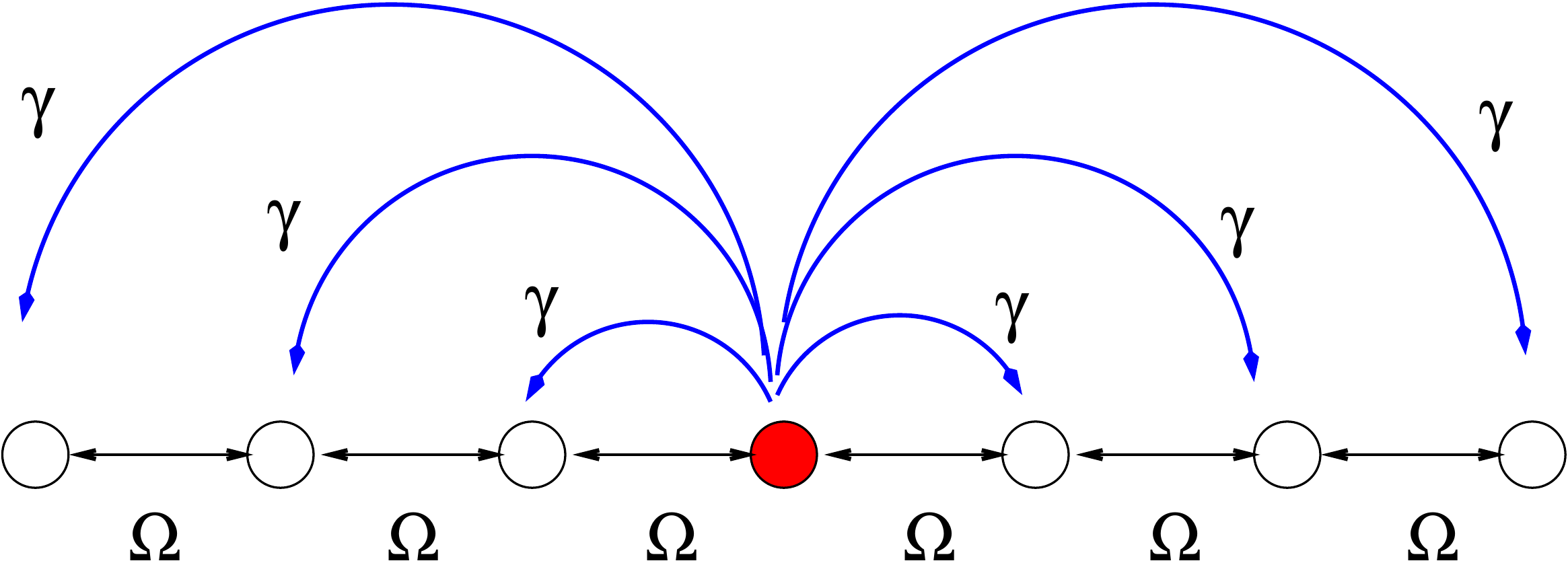}
\caption{Schematic representation of the  model 
  described by the  Hamiltonian (~\ref{H}). The particle  can hop
  through a one dimensional ring with disordered site energies,  in presence of
  nearest-neighbor tunnelling amplitudes, $\Omega$. $V_{LR}$ represents
  the long range (distant independent) hopping with strength $\gamma$. 
}
\label{Model}
\end{figure} 

In a recent publication~\cite{lea} by some of the Authors of the
present paper, it was found a common features of long-range
interacting systems, named {\it Cooperative Shielding}. 
This effect has been discussed  in a many-body spin
system in Ref.~\cite{lea}. There, it was shown
that Shielding is able explain many ``contradictory'' dynamical and transport features 
in systems with long-range interactions, as the ones mentioned above. 
Indeed, contrary to the common lore, which claims that
propagation of perturbation is very fast in long-range interacting systems,
it was found that  even in the regime of very large long-range interaction strength  there are subspaces where the evolution is determined by an 
emergent short ranged Hamiltonian. 

Here we analyze the Cooperative Shielding effect in a different model: 
a single excitation model of transport with long-range hopping.
We also discuss the  consequences of such effect on transport and
localization. 
Specifically, here we focus on models with an infinite interaction range, which
are  representative of the whole class of long-range interacting
systems~\cite{ruffo,adame}. Despite its apparent simplicity, 
infinite range hopping can be realized experimentally in ion
traps~\cite{iontrap} where linear spin chains have been recently
emulated with a spin-spin interaction decaying with the distance as
$1/r^{\alpha}$ with $0 \le \alpha \le 3$. The case $\alpha=0$  corresponds to an infinite
interaction range, which is discussed here. 
%Such case is also relevant to
%model excitonic transport 
%in engineered exciton wires~\cite{exciton} where it is induced by the
%coupling to a large wavelength cavity mode. 
Moreover, it is
routinely used to model superconductivity in 
ultra-small metallic grains~\cite{bcs} and non-equilibrium phenomena
around a phase transition in strongly correlated
materials~\cite{fabrizio}.

\section{The Model and the Energy Gap}

We  discuss the shielding effect by mean of  a paradigmatic model
for quantum transport, e.g.
a $1d$ Anderson model~\cite{Anderson} (described by $H_0$) with $N$ sites, 
with the addition of  long-range hopping terms $V_{LR}$, having a distance
independent coupling amplitude. The   Hamiltonian, see
also Fig.~\ref{Model}, is given by:
\begin{multline}
H = H_0+V_{LR}=H_{NN}+D+V_{LR} =\\
- \Omega \sum_{i} \left( | i \rangle
  \langle i+1| + h.c. \right) + \sum_i E^0_i |i\rangle \langle i| 
\ -\gamma \sum_{i\ne j} |i\rangle
\langle j| .
\label{H}
\end{multline}
The basis states $|j\rangle$ can represent the state of a particle
localized at the  site $j$ as it is usually assumed
 in tight-binding models of
electronic transport or, as in models of excitonic transport, 
 an excited two-level system at site $j$, 
when all the other two-level systems are  in their ground state. 
In Eq.~(\ref{H}), $H_{NN}$
represents the nearest-neighbor hopping with  $\Omega>0$ and
 $D$ the disordered part, since  
we assume $E^0_i$ to be random variables uniformly distributed in 
the interval $[-W/2,+W/2]$ ($W$ is the disorder strength).  
In this way $H_0=H_{NN}+D$ is exactly the Anderson Hamiltonian~\cite{Anderson}
while the long-range hopping with $\gamma>0$ is fully contained in   
 $V_{LR}$. Note that our main results are largely independent of the
 distribution of the site energies $E_i^0$. 
  
\begin{figure}[t!] 
\centering
\includegraphics[width=\columnwidth]{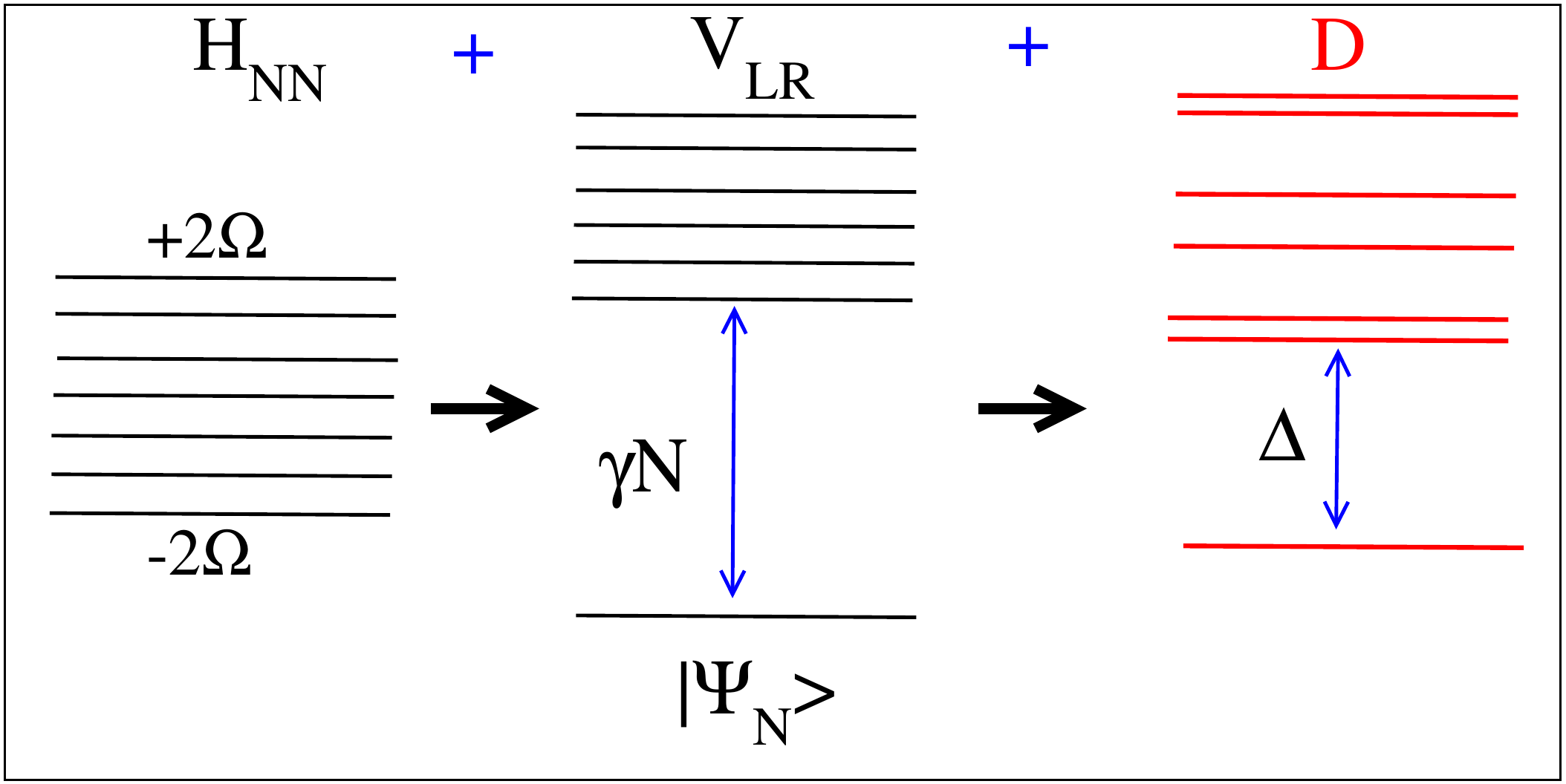}
\caption{Schematic representation of the energy levels of the system
  (~\ref{H}), as different terms are turned on. On the left column the spectrum of
   $H_{NN}$  only is shown. Middle column: creation of a gap once the long-range
  hopping term $V_{LR}$ is added. Right column: the
  effect of static disorder on the spectrum of the total
  Hamiltonian. Once disorder is added, the  gap $\Delta$  survives only for 
  $\gamma > \gamma_{cr}$, see Eq.~(\ref{gapcr}).
}
\label{Gap}
\end{figure} 
%{\it Physical relevance:}
%Shielding implies that the excited states are
%affected by disorder similarly to the states of the Anderson model,
%acquiring a localized nature despite the presence of long range hopping. 
%Moreover the ground state shows
%{\it Cooperative Robustness} to disorder, that is the 
% disorder strength needed to localize it, increases with the system size. On the
% other side, shielding implies that localization is possible for the
% excited states for a much weaker disorder, corresponding to the one
% of the $1D$ Anderson model. 
Let us mention two important 
models equivalent to Eq.~(\ref{H}). The first  is 
 the spin system: $$H=\sum_k h_k \sigma_k^z -2\Omega \sum [\sigma_x^k \sigma_x^{k+1} + \sigma_y^{k} 
\sigma_y^{k+1}]$$ $$ -2\gamma \sum [\sigma_x^l \sigma_x^m +
  \sigma_y^l \sigma_y^m]$$  in the single
excitation manifold, which
can be implemented in ion trap experiments~\cite{iontrap}. 
The second is 
the discrete BCS model~\cite{bcs}  (which is equivalent to
Eq.~(\ref{H}) in the limit of vanishing nearest-neighbor hopping),
 $$H_{BCS}= \sum_i E^0_i |i\rangle \langle i| 
\ -\gamma \sum_{i\ne j} |i\rangle \langle j|, $$ where 
$|i\rangle=|E_i^0 \uparrow; E_i^0 \downarrow \rangle$ is a Cooper pair state, 
where one electron occupies a single
particle state with energy $E^0_i$ and the other one is the time reversed state.
%Interestingly, current ion trap experiment with long range interaction could 
%then be used to analyze relevant features of superconducting grains.  

Let us first consider the case of no disorder $W=0$, so that $H=H_{NN}+V_{LR}$. 
The eigenvalues of  $H_{NN}$ in (~\ref{H}) can be computed
exactly and are given by,
$$ E^{NN}_q= -2 \Omega \cos{(2 \pi q/N)}$$
  with  $q=1,\ldots,N$, together
with  the components of the  eigenstates
\begin{equation}
\braket{k}{\psi_q}= e^{2 \pi i k q /N }/\sqrt{N}
\label{psiq0}
\end{equation}
in the site basis $|k\rangle$.
The ground state, corresponding to  $q=N$ and energy $ E^{NN}_{N}=-2 \Omega
$,
\begin{equation}
|\psi_N \rangle=\frac{1}{\sqrt{N}} \sum_{k=1}^N |k\rangle.
\label{SR}
\end{equation}
is fully symmetric and  extended in the site basis. 
This state is   also  eigenstate of
$V_{LR}$ with eigenvalue  $-(N-1) \gamma$. 
All the other eigenstates of $V_{LR}$ are degenerate with 
eigenvalue $+\gamma$, and can always be chosen to coincide with the other
eigenstates of $H_{NN}$, since $[H_{NN},V_{LR}]=0$. Note  that $V_{LR}$ has only
 two different eigenvalues: $-(N-1) \gamma$ which is not degenerate
 and $+\gamma$ which is $N-1$ degenerate. 
The eigenvalues of the Hamiltonian, $H_{NN}+V_{LR}$,
 can be written as,
$$
E_q=-2\Omega \cos{(2\pi q/N)} +\gamma -N\gamma \delta_{qN},
$$ 
with $q=1,...,N$. It follows that 
$H_{NN}+V_{LR}$ is characterized by a ground state  $|\psi_N\rangle$
separated by an energy gap $N\gamma$ from the excited states, see
Fig.~\ref{Gap} (central column).

Let us now analyze the case of $W\ne 0$. 
In this case $H_0=H_{NN}+D$  does not commute anymore with
$V_{LR}$ and disorder tends to mix the two eigensubspaces of
$V_{LR}$. To understand this point, 
it is convenient to write the
total Hamiltonian in the eigenbasis of $V_{LR}$, see Eq.~(\ref{psiq0}),
which diagonalize $H_{NN}+V_{LR}$. 
In this basis $D$ becomes a full matrix and  the variance  of the matrix
elements $D_{qq'}$ which connects the eigenstates of $V_{LR}$, see
Eq.~(\ref{psiq0}),  can be easily computed~\cite{adame} as $\langle |D_{qq'}|^2
\rangle = \frac{W^2}{12 N}$, where $\langle...\rangle$ stands for the
average over disorder.  
%Thus the typical strength of
%interaction  between the states $|\psi_q \rangle$ is  $\simeq \frac{W}{\sqrt{12 N}}$.

Concerning the energy gap, one could expect that,
for large $N \gamma$  or weak disorder strength $W$,
the mixing between the two eigensubspace of $V_{LR}$ is also
weak and an energy gap will be  present in the spectrum
of the total Hamiltonian, too. 
Indeed, the energy gap can be computed exactly in the limit $W \gg \Omega$ and
large $N$ and it coincides with the Richardson solution
of  the superconducting gap~\cite{bcs}: 
\begin{equation}
\Delta=W/\sinh{(W/ N \gamma)},
\label{Delta}
\end{equation}
 from which we have that the gap increases with the system size,
 indeed $\Delta \approx 2W e^{-W/N\gamma}$ for $W \gg N\gamma$
and $\Delta \approx N\gamma$ for $N\gamma \gg W$. 
For any finite size system, 
we can talk about an energy gap only when the energy distance
between the ground state and the first excited state is larger than
the mean level spacing $\delta$ (average distance between the energy levels).
Thus,  the condition for the destruction of the gap $\Delta$ can be 
estimated from $\delta \simeq \Delta$.
 For $\gamma=0 $ and $ W \gg \Omega$ one has $\delta \simeq W/N$.
From the condition $\Delta/\delta \simeq 1$, 
 we can find a critical $\gamma$, below which the gap disappears:
\begin{equation}
\gamma_{cr} \simeq \frac{W}{N \operatorname{arsinh} (N)} \approx \frac{W}{N \ln{2N}}
\quad \mbox{for} \quad N\gg 1.
\label{gapcr}
\end{equation}

\begin{figure}[t!] 
\centering
\includegraphics[width=\columnwidth]{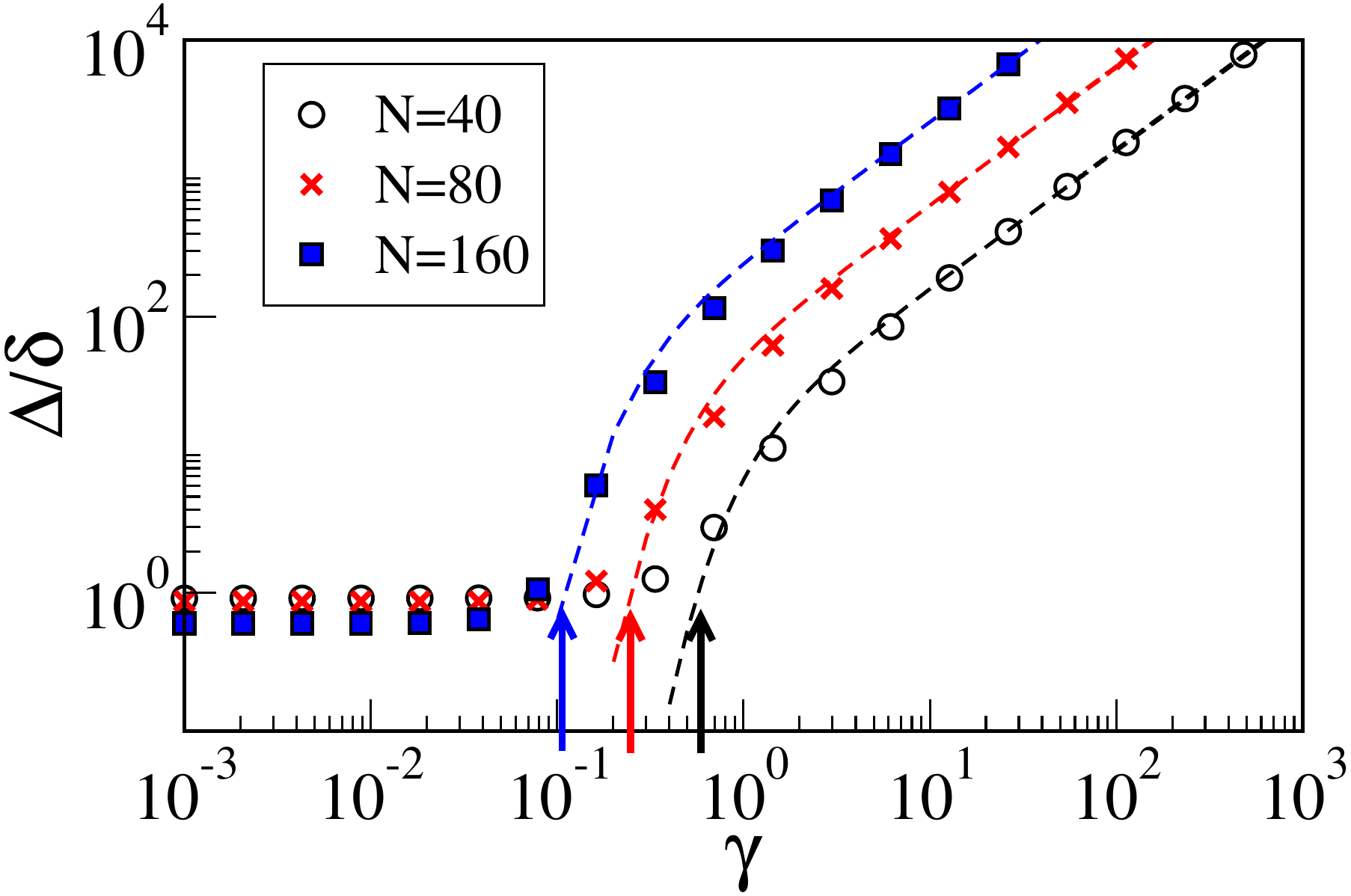}
\caption{
Rescaled gap $\Delta/\delta$ with $\delta = W/N$, average spacing between energy
levels, as a function of the long-range hopping strength $\gamma$ 
for different $N$ values as indicated in the caption.
Here is $W=100$, $\Omega = 1$. Vertical arrows indicate the critical 
values $\gamma_{cr}$ given in  Eq.~(\ref{gapcr}). Dashed lines represent
the Richardson solution as given in   Eq.~(\ref{Delta}).
}
\label{taugap3}
\end{figure}

In Fig.~\ref{taugap3} we show the critical strength $\gamma_{cr}$ as vertical 
arrows together with the analytical estimate given in Eq.~(\ref{Delta}). As 
one can see the ratio $\Delta/\delta$ is well described by Eq.~(\ref{Delta})
for $\gamma > \gamma_{cr}$ (compare symbols with dashed lines).

\section{Cooperative Shielding}

In this Section we want to analyze the Cooperative Shielding effect both
without disorder, where it is exact, and with disorder, where it is
effective for a finite time scale, diverging with the system size. 

As we have shown in the previous Section, $V_{LR}$ has only two
eigenvalues, one corresponding to the fully symmetric state,
Eq.~(\ref{SR}), and the other one corresponding to a $N-1$ degenerate subspace.

For $W=0$ we have that  $[H_{NN},V_{LR}]=0$ and the excited states of
the Hamiltonian $H_{NN}+V_{LR}$ coincides with the degenerate subspace
of $V_{LR}$. Here we show that in the excited states subspace of
$H_{NN}+V_{LR}$, the dynamics occurs as if long range would be
absent and it is determined by $H_{NN}$ only. We will show this point in two different ways. 

First, let us consider a trivial and general case, i.e. a system described by
$H=H_0+V$ with  $[H_0,V]=0$. We also assume that the spectrum 
$V$ is degenerate in one of
its eigensubspaces ${\cal V}$, so that $V|v_k\rangle=v |v_k\rangle$
$\forall |v_k\rangle \in {\cal V}$. 
The evolution  of any initial state $|\psi(0)\rangle$
belonging to ${\cal V}$ is simply given by: 
$|\psi(t)\rangle= e^{-ivt/\hbar} e^{-iH_0t/\hbar}
|\psi(0)\rangle$. Since the only effect of $V$ is to induce a global
phase which has no effect on any observables, we can say that  the dynamics, starting from
an eigensubspace of $V$, is shielded from $V$ and determined only by $H_0$. 
%Note that it makes sense to speak about dynamics since the shielded
%eigensubspace has a high dimensionality due to its degeneracy. 
%t is also clear that if the initial state has large components in more than
%one eigensubspace of $V$, the dynamics will not be shielded. 
From the above discussion it follows that for $W=0$, shielding in the excited states subspace, which coincides
with the eigenspace of the degenerate eigenstates of $V_{LR}$, is
exact  for any value of $\gamma$ and $N$.
In the following we will show that, even when $[H_0,V_{LR}] \ne 0$, 
shielding can still persist up to a very long
time, diverging with the system size (cooperativity). 
This is rather counterintuitive, since if $[H_0,V_{LR}] \ne 0$, $H_0$ mixes the
 different eigensubspaces of $V_{LR}$ so that one might  expect
 a dynamics strongly dependent on $V_{LR}$ itself.

We can show the shielding effect in the case  of no disorder also
considering the dynamics.
 For the sake of simplicity let us  add a term $-\gamma \sum_i
|i\rangle \langle i|$ to the Hamiltonian $H_{NN}+V_{LR}$. Such a
choice corresponds to a constant energy shift
equal to $-\gamma$ of the site energies and  does not affect the
dynamics of the system. 
%Including this term in $V_{LR}$, implies that
%only one of its eigenvalues, corresponding to the  state $|\psi_N
%\rangle$ , is non-zero   and equal to $-N \gamma$. 
%This eigenstate also corresponds to the
%ground state of $H_{NN}$, given in Eq.~(\ref{SR}).
%All the other  $N-1$ eigenstates of $V_{LR}$ are degenerate with null
%eigenvalue. 
Let us consider a generic initial state, $|\psi \rangle= \sum_k c_k |k\rangle$,
written on the site basis $|k\rangle$. Evolution is given by  the Schrodinger equation
with Hamiltonian,  $H_{NN}+V_{LR}-\gamma \sum_i
|i\rangle \langle i|$, and we have:
\beq
i \hbar \frac{dc_k}{dt}= -\Omega (c_{k-1}+c_{k+1}) -\gamma \sum_j c_j
\label{em}
\eeq
where the first term on the r.h.s. represents  the nearest-neighbor evolution
and the second is due to long-range hopping.
Equations of motion can be rewritten in terms of the macroscopic
quantity :
$$
Z (t)=\frac{1}{\sqrt{N}}\sum_k c_k (t),
$$
namely 
\begin{equation}
\left\{
\begin{array}{cl}
 i \hbar \frac{dc_k}{dt}= -\Omega (c_{k-1}+c_{k+1}) -\gamma \sqrt{N} Z  \\
\\
i \hbar \frac{d Z}{dt}= (-2 \Omega  -\gamma N)  Z   \,
     \end{array}
\right.
\label{em2}
\end{equation}
Note that for the ground state $|\psi_N \rangle$, see Eq.~(\ref{SR}), we have $Z=1$ while
for all the other eigenstates $|\psi_q\rangle$ with $q=1,...,N-1$ we
have $Z=0$. 
 
From the second of Eq.~(\ref{em2})  it follows trivially  that if  $Z(0)=0$ then $Z(t)=0$ for all times.  
 Thus the evolution
of any initial  state orthogonal to the ground state (characterized by $Z=0$), namely
any combination of excited states,  will be determined only by the nearest-neighbor
part, as if the long-range interaction would be absent, see first
equation in Eq.~(\ref{em2}). 
The absence of long-range hopping in the $Z=0$ subspace for $W=0$ has
some counter-intuitive effects. 
For instance in absence of nearest-neighbor interaction if we start
with a particle in an antisymmetric superposition on two sites, so
that $Z=0$, we have that the particle will not go anywhere, despite
the  distance independent tunneling hopping
amplitude which connects all the sites. 
%This effect has been named 
% self trapping~\cite{selftrapping} in literature, and it can be seen as
% consequence of the shielding effect discussed here.  

Note that, for the case of no disorder, 
{\it shielding} from long-range interaction  involves a
subspace of dimension $N-1$  (all the eigenstates orthogonal to
the ground state $|\psi_N\rangle $ of $H_{NN}+V_{LR}$ satisfy the
condition $Z=0$) thus including  almost the whole Hilbert space.  

The question we want to address now if whether this effect survives in
presence of disorder, which breaks the symmetry of the system so that
the $V_{LR}$ does not commute anymore with the rest of the
Hamiltonian.
Disorder also tends to mix the 
two eigensubspaces of $V_{LR}$ and this
could lead, in general,  to the destruction of shielding in the
excited states. Indeed, in presence of disorder, the excited states do
not lie completely in the degenerate subspace of $V_{LR}$, so that a
random initial state in the excited state subspace will in general
have components in more than an eigensubspace of $V_{LR}$ and thus
$V_{LR}$ can be relevant for its dynamics.
Nevertheless, even for $W \ne  0$ and in the regime of large gap
$\gamma \gg \gamma_{cr}$,  the  mixing
between the two  eigensubspaces of $V_{LR}$  induced
by disorder, is weak. 
Under these conditions, any initial state in the
excited subspace of $H$, will mainly lie in the degenerate subspace of
$V_{LR}$ so that it will be dynamically  shielded from the long-range
term for a very long time. 
Interestingly, on increasing the system size, the
gap, see Eq.~(\ref{Delta}), increases and shielding becomes stronger as it
will be shown here below.

 \begin{figure}[t!] 
\centering
\includegraphics[width=\columnwidth]{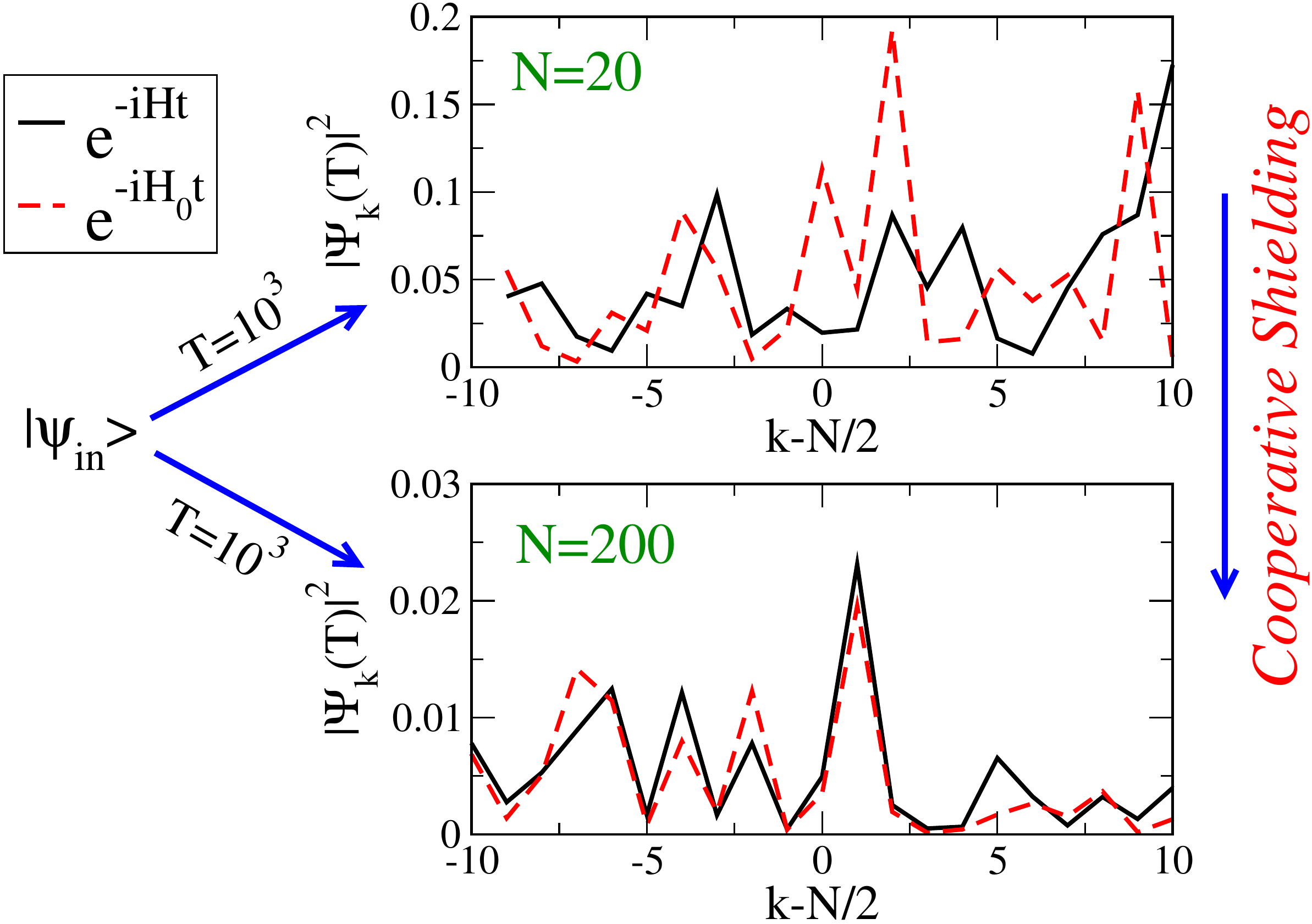}
\caption{Probability  to be on the site   $k$-th,  at the time $T=10^3$ for $N=20$
  (upper panel) and $N=200$ (lower panel). In both panels, full black
  lines have been obtained by evolving the initial state $|\psi_{in}\rangle$ with the full
  Hamiltonian $H$, while the red dashed curves are obtained with $H_0$, see
  Eq.~(\ref{H}). The initial state $|\psi_{in}\rangle$ has be chosen to be  a random superposition
  of the excited states of $H$. 
Other data are: $\Omega=1, W=1, \gamma=1$ and we are in the regime
$\gamma>\gamma_{cr}$, see Eq.~(\ref{gapcr}).
}
\label{csH0}
\end{figure} 

\begin{figure}[t!] 
\centering
\includegraphics[width=\columnwidth]{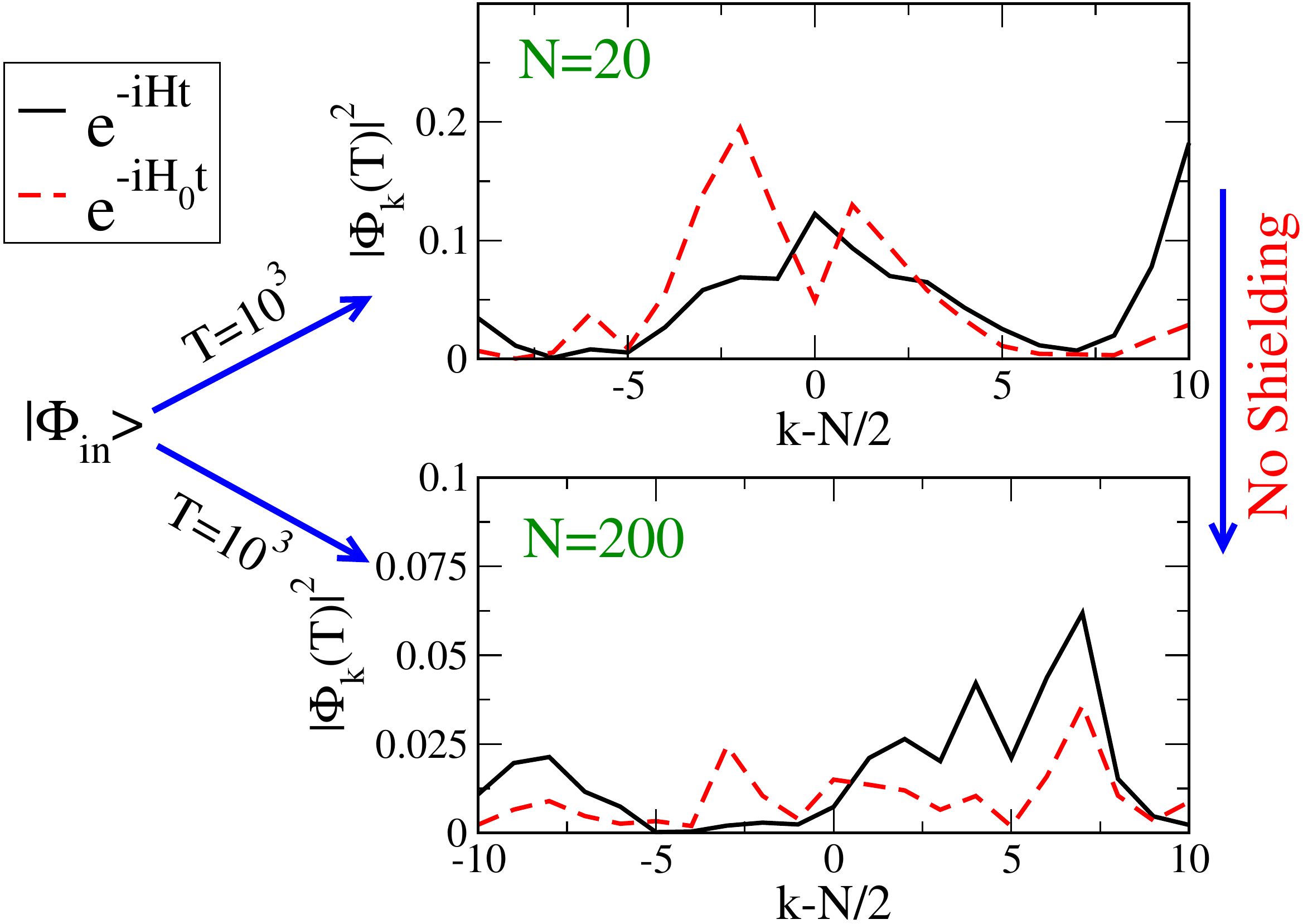}
\caption{Probability  to be on the site   $k$-th,  at the time $T=10^3$ for $N=20$
  (upper panel) and $N=200$ (lower panel). In both panels, full black
  lines have been obtained by evolving the initial state $|\Phi_{in}\rangle$ with the full
  Hamiltonian $H$, while the red dashed curves are obtained with $H_0$, see
  Eq.~(\ref{H}). The initial state $|\Phi_{in}\rangle$ has be chosen to be  a superposition
  of the excited states and the ground state of $H$, see text. 
Other data are: $\Omega=1, W=1, \gamma=1$ and we are in the regime
$\gamma>\gamma_{cr}$, see Eq.~(\ref{gapcr}).
}
\label{csH1}
\end{figure} 

\subsection{A qualitative dynamical analysis}

In this Section we will give a qualitative analysis
of the Shielding effect
based on the dynamics. 
We choose as initial state $|\psi_{in}\rangle$ a 
random superposition of the  excited  states of $H$ and
we compare its time
evolution (in the regime $\gamma \gg \gamma_{cr}$)
% belonging to the
%degenerate eigensubspace of $V_{LR}$ (for simplicity we took
%$|\psi_0\rangle= \frac{1}{\sqrt{2}} (|N/2\rangle -|N/2-1\rangle)$, 
under both the full
Hamiltonian $H$ and the Anderson Hamiltonian $H_0$ (which
does not contain  the long-range term). In order to have a consistent
spreading of the initial wave packet we chose a small value of
disorder $W/\Omega=1$. 
%From the discussion above, since $\gamma \gg \gamma_{cr}$
%such state lies mainly in the excited states subspace of $H$. 
In Fig.~\ref{csH0} we plot the  probability distribution in the site
basis at a fixed time $T$ for two different $N$ values. While for
small $N$  the unperturbed and full dynamics give completely different
results (upper panel), for a sufficiently large $N$ value the two
distributions are close one to each other (lower panel) showing how on
increasing $N$ (cooperativity) at fixed time $T$ the full dynamics occurs as if long-range
hopping would be absent (shielding).  
This simple example  shows that the presence of a gap (the condition  $\gamma> \gamma_{cr}$ is verified in both panels of Fig.~\ref{csH0})
is not enough to have shielding, which can thus be  defined 
only within a suitable time scale. This point will be discussed in the
next subsection.

Note that  in the regime of large gap shielding is effective only for
an initial state which lies in the excited state subspace. 
On the other hand, for initial states with a strong component on the
ground state, shielding will not  be effective. Indeed in this case
the initial state will have large components on eigenstates of $V_{LR}$ with
different eigenvalues, so that $V_{LR}$ can impact the dynamics. 
To illustrate this point, let us 
consider another initial condition $|\Phi_{in}\rangle= 1/ \sqrt{2n+1}
\sum_{k=-n}^{k=n} |k\rangle$ with $n=4$ which has  strong components
on both eigensubspaces of $V_{LR}$. 
As one can see from Fig.~\ref{csH1}, as $N$
increases the shielding effect does not occur, contrary to the other
initial conditions considered in Fig.~\ref{csH0} for the same
parameters.

\subsection{Time scale of fidelity decay}

\begin{figure}[t!] 
\centering
\includegraphics[width=\columnwidth]{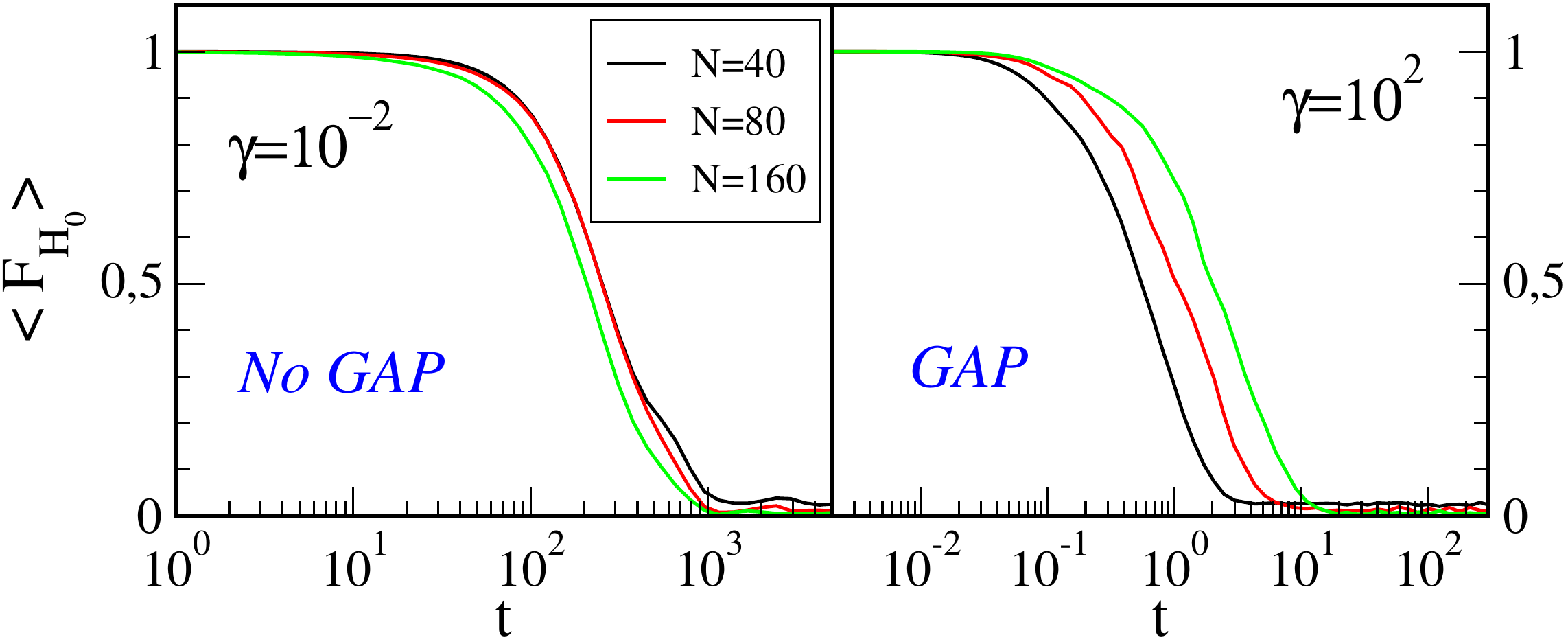}
\caption{Average  
Fidelity {\it vs} time starting from a random
  superposition of the $N-1$ excited states and different $N$ values
as indicated in the legend.
Left  panel:  small coupling $\gamma=10^{-2} < \gamma_{cr}$;
right panel : large coupling $\gamma=10^{2} > \gamma_{cr}$.
Other data are : $\Omega=1, W=100$. }
\label{ffb}
\end{figure}

To analyze in a more quantitative way the Shielding effect, we consider the  Fidelity, defined as the  
overlap probability   between a state evolved with the
full Hamiltonian (\ref{H}) and that evolved with 
$H_0=H_{NN}+D$ (without the long-range term). As initial condition we
choose a state $|\psi_{in}\rangle$ composed by a random superposition of all the excited
states of $H$, and we compute the Fidelity as:
\begin{equation}
F_{H_0}(t)=|\langle \psi_{in}| e^{i H_0 t/\hbar}e^{-i  Ht/\hbar} |\psi_{in}\rangle|^2.
\label{F}
\end{equation}
Examples of Fidelity decay in time are reported 
in Fig.~\ref{ffb}. 

In the two panels of Fig.~\ref{ffb} the behaviour of the
average Fidelity, see Eq.~(\ref{F}), $\langle F_{H_0} \rangle$, taken
over different
random realizations
 {\it vs} time is shown in two different regimes (above and
below $\gamma_{cr}$), see Eq.~(\ref{gapcr}). 
As one can see, while for
$\gamma \ll \gamma_{cr}$ (gapless region) 
the Fidelity decay is almost independent of the system size, 
the shielding effect manifests itself for $\gamma \gg \gamma_{cr}$ (gapped region)
where the Fidelity increases with the system size.

\begin{figure}[t!] 
\centering
\includegraphics[width=\columnwidth]{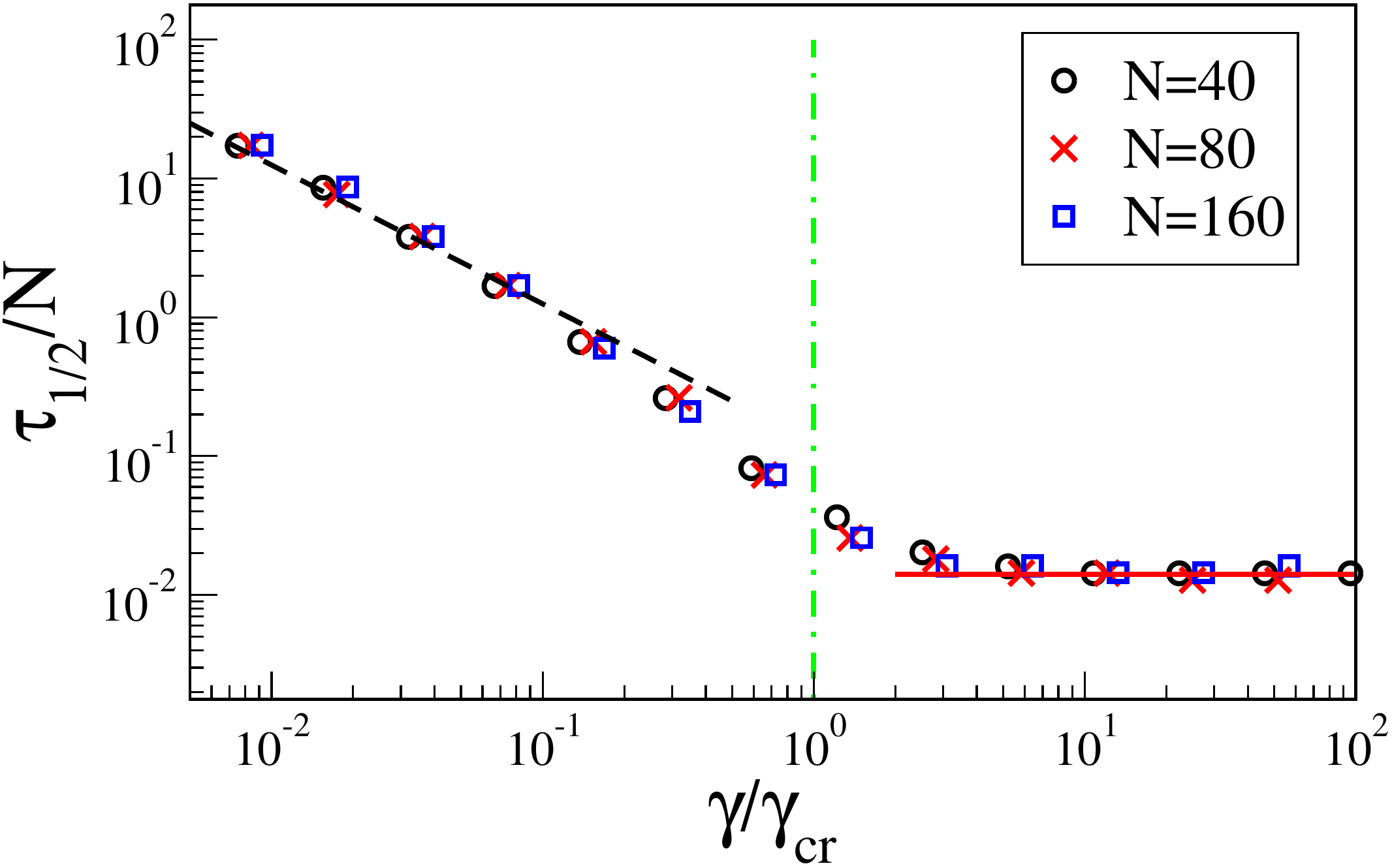}
\caption{Rescaled half time $\tau_{1/2}/N$ {\it vs}
  $\gamma/\gamma_{cr}$ for different $N$ values as indicated in the
  legend. As initial state 
a random
  superposition of the $N-1$ excited states has been chosen. 
Other data are : $\Omega=1, W=100$. 
Horizontal  full line 
represents the theoretical
prediction,  $\tau_{1/2}/N = c_2/W$ in the gapped regime. The dashed line
stands for $c_1\gamma_{cr}/\gamma$. $c_{1,2}$ are  fitting constants. 
The vertical dot-dashed line is $\gamma=\gamma_{cr}$.}
\label{ff}
\end{figure} 
%In the gapped regime it is also possible to estimate the dependence of
%$\tau_{1/2}$ on $N$ (see Supplementary
%Material) and we have $\tau_{1/2} \simeq N/W$.

If shielding would be perfect, Fidelity
would remain equal to one for all times. Thus, to quantify shielding we use
the time $\tau_{1/2}$ at which
the average (over disorder) Fidelity decreases to
one half. This ``shielding time'' $\tau_{1/2}$ rescaled to the system size
  is plotted in  Fig.~\ref{ff} {\it vs} $\gamma/\gamma_{cr}$ for different
system sizes. For a fixed $W$, all data collapse on a single curve given by,
\begin{equation}
\frac{\tau_{1/2}}{N} = \left\{
\begin{array}{lll}
\displaystyle c_1 \frac{\gamma_{cr}}{\gamma} \quad &{\rm for} \quad 
&\displaystyle   \gamma \ll \gamma_{cr}\\
      & \\
\displaystyle   \frac{c_2}{W} \quad &{\rm for} \quad 
&\displaystyle \gamma \gg \gamma_{cr} \\   
     \end{array}
\right.
\label{tau12}
\end{equation}
where $c_{1,2}$ are fitting constants. From Eq.~(\ref{tau12}) it
follows that for $\gamma \ll \gamma_{cr}$, one has $\tau_{1/2} \propto
W/(\gamma \ln(2N))$. 
Since $\gamma$ represents the strength of the perturbation, it is 
expected that Fidelity decays faster as $\gamma$ increases. 
Note also that
$\tau_{1/2}$ decreases with $N$  in this regime.  
On the other side
when $\gamma \gg \gamma_{cr}$ ({\it gapped} regime), we have $\tau_{1/2}
\propto N/W$, which means  that the shielding time becomes completely independent of
the long-range coupling strength
$\gamma$ ({\it shielding}) and it increases with the system size ({\it
  cooperativity}). The expression  $\tau_{1/2} \simeq
N/W$ in the gapped regime  can also be obtained  analytically (see Appendix B). Let us stress 
that this behaviour is peculiar of the homogeneous long-range
hopping. Indeed if one considers the case of random long-range hopping, namely a random coupling between different sites, there is no gap in the spectrum of the Hamiltonian and $\tau_{1/2}$ always decreases with $\gamma$ as expected (see Appendix B).

\section{Consequences of Shielding on Propagation of Perturbations}
Here we show the consequences of the shielding effect on the velocity
of propagation of perturbations, which is 
a major topic of investigation in recent literature. In particular one
of the main question is to understand whether the propagation of
excitations in systems with long-range interaction remains or not
confined to an effective light
cone~\cite{Hastings2006,Schachenmayer2013,Eisert2013,iontrap,Gong2014,Mazza2014,Metivier2014,Feig2015,Storch2015},
as defined by the Lieb-Robinson bound~\cite{Lieb1972} and its
generalizations~( see~\cite{Storch2015} and references therein). Indeed in 
systems with nearest-neighbor interaction, it has been
proved~\cite{Lieb1972} that the propagation of perturbations has a
finite velocity proportional to the nearest-neighbor coupling strength
and independent of the system size, so that it defines an effective
"light-cone". Outside such  light-cone, the propagation is
exponentially suppressed. 
On the other hand, in recent experiments with trapped ions, it was
observed that for long-range interaction, the light-cone picture is no
longer valid, the dynamics becomes nonlocal and perturbations propagate
non-linearly in time. 

This subject has been recently discussed in Ref.~\cite{lea} by some of
the authors of this paper. There it is shown that, due to {\it
  Cooperative Shielding}, together with very
fast spreading of perturbations, long-range interacting system can
also display freezing or  effective short-ranged dynamics
(constrained within a light-cone). This occurs if the initial
conditions belong to a shielded subspace, and it is valid for a finite
time, which increases with the system size.  

\begin{figure}[t!] 
\centering
\includegraphics[width=\columnwidth]{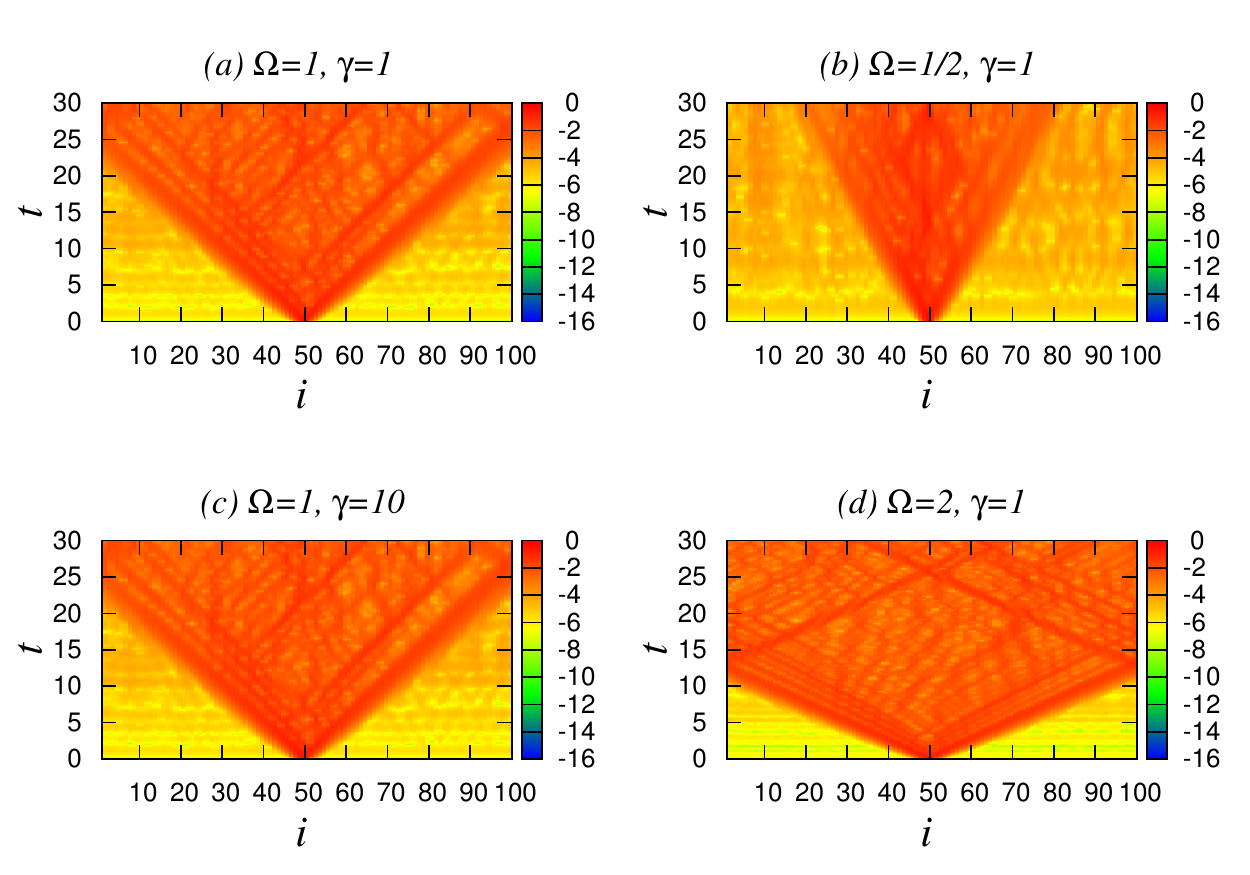}
\caption{Probability (in log scale) to be on site $i$ at time $t$. The
  dynamics has been computed starting from the initial state
  $|\psi_0\rangle= (|N/2\rangle -|N/2-1\rangle)/\sqrt{2}$. 
Common parameters for all panels are $N=100,W=1$.
In panels $a,c$ we change $\gamma$ keeping all other parameters fixed, showing
that the velocity of propagation of the initial perturbation is
independent of $\gamma$ (long-range coupling strength) within the time scale in the figure. In
panels $b,c$ only $\Omega$ (nearest-neighbor hopping strength)
 is changed, keeping all other parameter
fixed, showing that the velocity of propagation is proportional to $\Omega$. 
}
\label{prop}
\end{figure}

Here we analyze the propagation of perturbations in our transport
model. Specifically we start from an excitation shared
anti-symmetrically between two nearest neighbor  sites: $|\psi_0\rangle=
(|N/2\rangle -|N/2-1\rangle)/\sqrt{2}$. Note that such state belongs to
the degenerate eigensubspace of $V_{LR}$ with $Z=0$. We let the state
evolve with the full Hamiltonian and compute the probability to be at
time $t$ on site $i$ along the chain. The results are shown in
Fig.~\ref{prop}. In panels $a,c$ we change only the long-range coupling
strength $\gamma$, while in panels $b,d$ we change only the
nearest-neighbor coupling $\Omega$, keeping all parameters fixed.
As one can see, the velocity of propagation describes a linear light-cone in all cases,
up to the time scale given in the figure, despite the presence of long-range coupling. In particular the shielding effect has the striking
consequence that the velocity of propagation is proportional to
$\Omega$, but completely unaffected by increasing $\gamma$. 

\begin{figure}[t!] 
\centering
\includegraphics[width=\columnwidth]{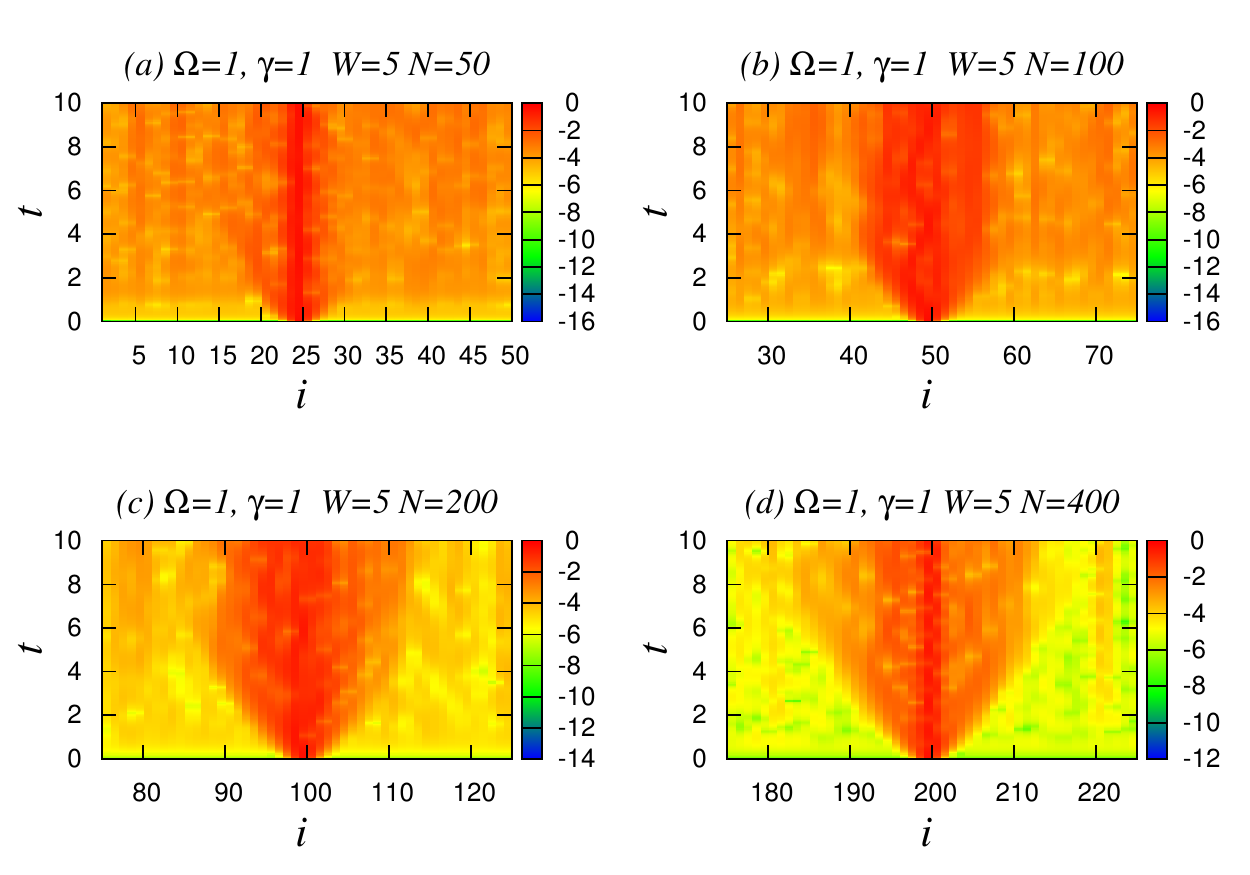}
\caption{Probability (in log scale) to be on site $i$ at time $t$. The
  dynamics has been computed starting from the initial state
  $|\psi_0\rangle= (|N/2\rangle -|N/2-1\rangle)/\sqrt{2}$. 
Common parameters for all panels are $\Omega=1,\gamma=1,W=5$.
As the number of sites increases, linear propagation of perturbations
holds for longer times, confirming the cooperative nature of the
shielding effect. 
}
\label{coopN}
\end{figure} 
In order to confirm the cooperative nature of the shielding effect, in
Fig.~\ref{coopN} we analyze the spreading of perturbations as the
system size is varied, keeping all other parameter fixed. As one can
see, as the system size increases, the linear propagation of
perturbations holds for longer times.

The results shown in this Section clarify the consequence of the shielding
effect on the spreading of perturbations:  due to shielding, even in
presence of a large long-range coupling strength, the spreading of
perturbation can be linear, as in the case of a short-ranged
dynamics. We have also shown that the time for which the propagation is linear increases with
the system size. 
On the other side, for different
initial conditions, shielding is no longer effective, as it was shown
in Fig.~\ref{csH1}, and the spreading
of perturbations can be very fast~\cite{iontrap}. In this sense shielding is able to
explain the contradictory behaviour of  long-range interacting
systems.

\section{Consequences of shielding on localization}
%Here we analyze the transport properties, focusing on 
%Anderson localization~\cite{Anderson}. 
In 1d tight-binding models with 
short-range interaction the addition of disorder induces localized
eigenstates thus suppressing transport~\cite{Anderson}. 
%The (finite size) critical disorder at which
%the vast majority of states localizes is given by $W_{AT} \approx
%10/\sqrt{N}$~\cite{Giulio}. 
It is also a common lore that
the presence of long-range hopping terms destroys localization~\cite{levitov}.
The following argument, for any dimension, had been usually advocated:
a particle in a specific site can tunnel 
to a distance $R$ under the resonant
condition $V \ge \Delta E$, where $\Delta E$  is the
energy difference between the two sites and $V=\gamma/R^{\alpha}$ is the tunnelling
coupling between them. 
Since the site energies are uniformly distributed between  $-W/2$ and
$W/2$, the probability to satisfy  the resonance condition is  
$\approx  V/W$. 
The number of  sites enclosed in two spheres of radius $R$ and $2R$ 
is proportional to
$R^d$, where $d$ is the embedding dimension of the lattice network, so that 
the number of resonant sites
$ N_{res} \propto V R^d /W\propto R^{d-\alpha}$. 
We can conclude that for short-range hopping
$\alpha>d$, $N_{res}$ goes to zero with the distance $R$,
while for long-range hopping $\alpha<d$ it
diverges with $R$. This general argument generates the
expectation that localization can occur only for short-range hopping. 

\begin{figure}[t!] 
\centering
\includegraphics[width=\columnwidth]{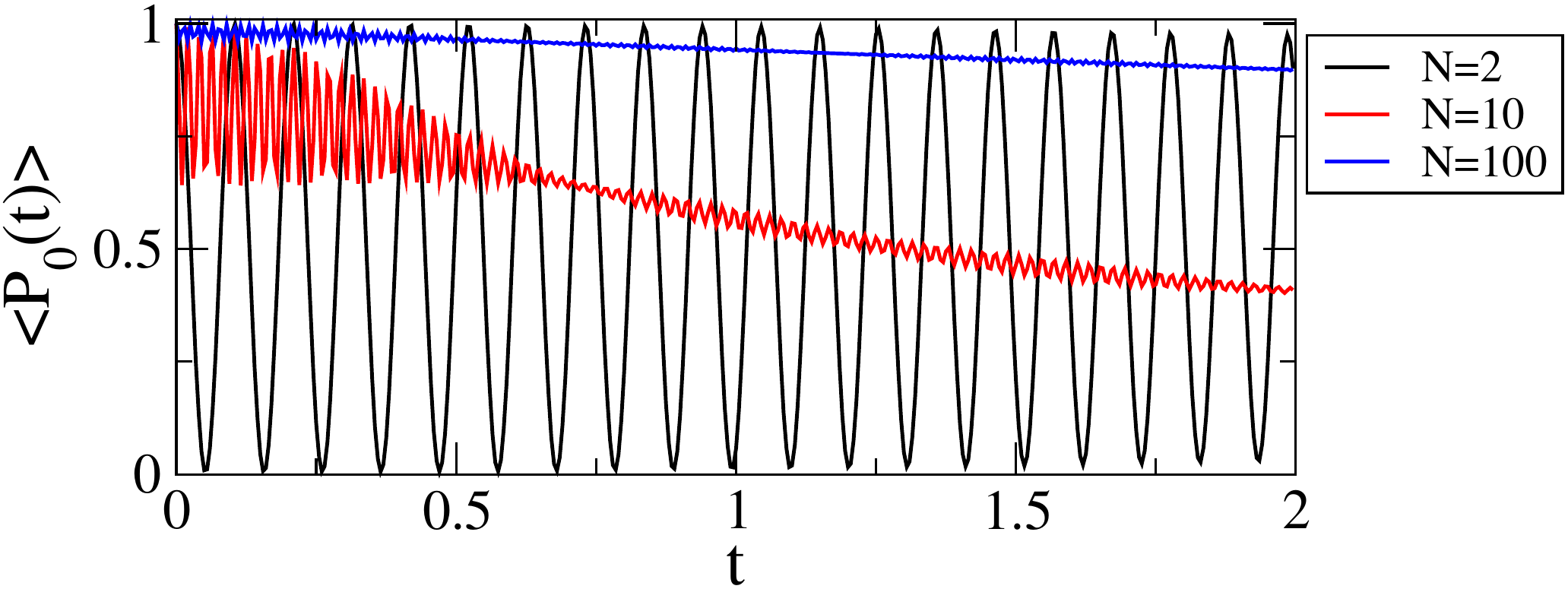}
\caption{Average survival probability {\it vs} time for different
  system sizes, see figure caption. At time $t=0$ only one site is
  occupied. Here is $\gamma=30 \gg W \gg \Omega$ so
  that the occupied site is at resonance with all the other
  sites. Nevertheless as the system size increases, the probability to
  find the system in the initial site increases.
}
\label{P0}
\end{figure}

For random long-range hopping 
this argument agrees with the results reported in literature, see
Ref.~\cite{levitov} and Appendix C.
On the other hand for homogeneous long-range hopping, which is our 
case, the situation is different: even if we have an infinite number
of resonances, shielding induces localization in the excited states
subspace. In order to show this point, let us consider an initial
state in which only one site is occupied
$|\psi_0\rangle=|1\rangle$. We consider the case in which $W/\Omega
\gg 1$, so
that without long-range the system would be highly localized and the
excitation would not spread much on the other sites in time. We also
set $\gamma \gg W$ so that the initial site is almost at resonance with
all the other sites. Let us compute the average over disorder survival
probability $\langle P_0(t) \rangle$, where $P_0(t)=|\langle
\psi_0|\psi(t)\rangle|^2$. In Fig.~\ref{P0} the average survival
probability is plotted {\it vs} time for different system sizes. For
$N=2$, site $|1\rangle$ is at resonance with site $|2\rangle$, so that
$P_0(t)$ oscillates between the two sites.
On  increasing the system size, one might expect  $P_0(t)$ to
decrease faster with the number of sites. Indeed the long-range
coupling is independent of the distance and site $|1\rangle$ is at
resonance with all other sites. Contrary to this expectation,
the probability to leave the initial site {\it decreases} with the number of
site. The results presented in Fig.~\ref{P0} show that localization
is enhanced on increasing the system size. How can we relate such
results with the Cooperative Shielding effect? In the large $N$ limit,
the ground state of the system is well approximated by the state
$|\psi_N\rangle$, Eq.~(\ref{SR}), which represents a symmetric superposition of
a  particle on all sites. An initial state localized only on one site
has a probability $1/N$ to be on the ground state and a probability
$(N-1)/N$ to be in the excited states, so that in the large $N$ limit,
a localized initial condition lies mainly in the excited states
subspace. As it has been shown in the previous Sections,  the excited states are shielded from long-range
interaction and the dynamics in such subspace is mainly determined by
$H_0$,  so that localization becomes possible. 
Again, both shielding
and cooperativity are shown by these results. 

In order to elucidate this point in a more quantitative way, we  study 
the participation ratio~\cite{pr} 
\begin{equation}
 PR= \left\langle 1/\sum_i |\langle i| \psi \rangle|^4 
\right\rangle, 
\label{PR}
\end{equation}
of the eigenstates $|\psi \rangle$  of the Hamiltonian   
(\ref{H}),
where  $\langle {\dots} \rangle$  stands for the ensemble average over 
different realizations of the static disorder.
For extended states, it increases proportionally to the system size, $N$,
while, for localized states, it is independent of $N$.
 
\begin{figure}[t!]
\centering 
\includegraphics[width=\columnwidth]{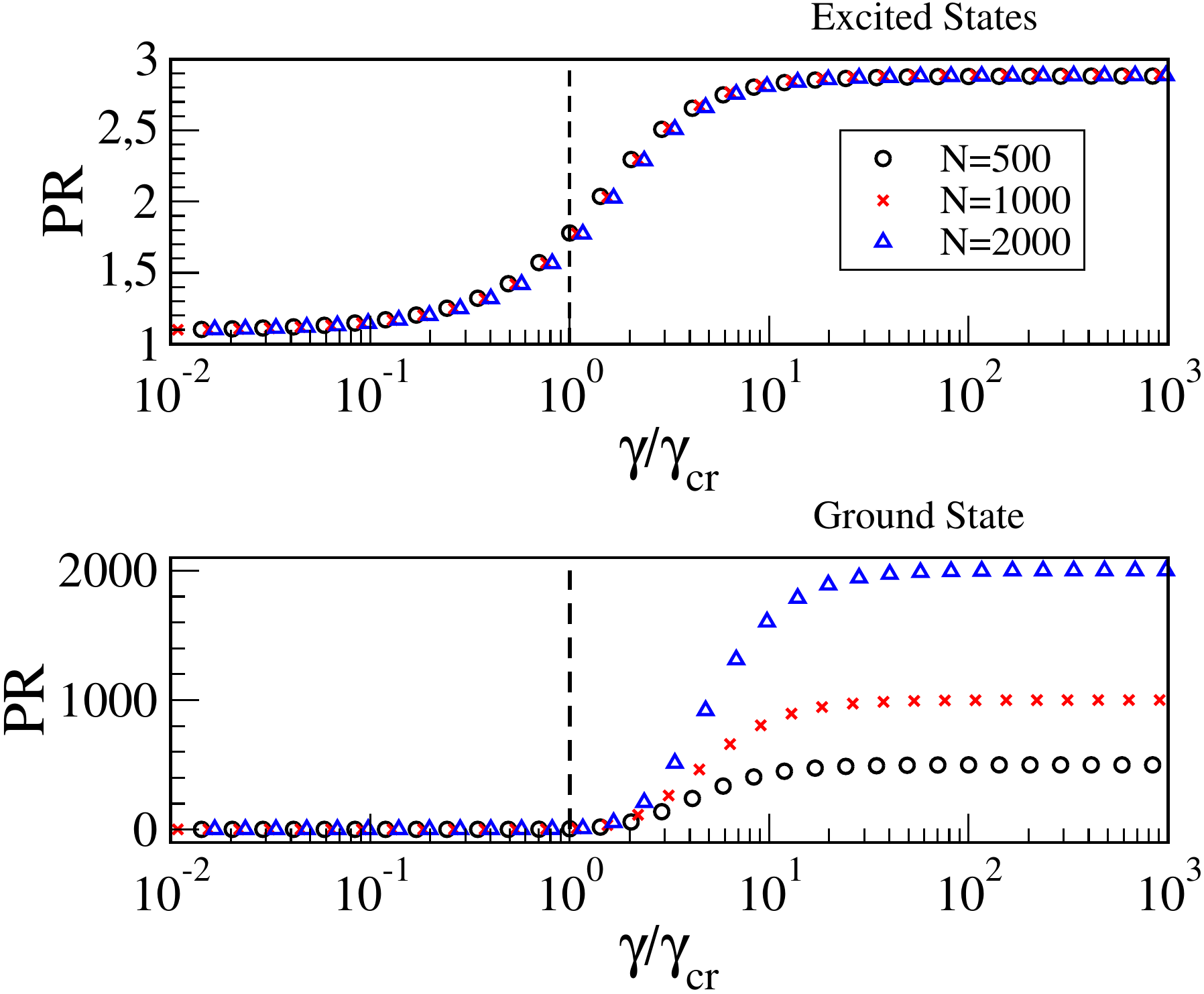}
\caption{Upper panel : average participation ratio
over the excited states 
 {\it vs} $\gamma/\gamma_{cr}$
and different system size.
Lower panel : Participation ratio  for the ground state.
Vertical dashed lines in both panels stand for 
$\gamma=\gamma_{cr}$. 
Other parameters  are: $\Omega=1, W=100$. 
}
\label{PRalpha}
\end{figure} 

In  Fig.~\ref{PRalpha} upper panel, the average PR of all the excited states 
 is shown as a function of the rescaled long-range hopping strength $\gamma/\gamma_{cr}$
for different system sizes. As one can see all points almost collapse in  a single curve,
showing that the PR depends only on  $\gamma/\gamma_{cr}$. Moreover,
above $\gamma_{cr}$, the PR reaches a constant value, which means that
it also becomes independent of $N$. This result shows that the excited states 
localize above $\gamma_{cr}$, Eq.~(\ref{gapcr}), for which the
gap in the spectrum opens and shielding becomes effective.  
This is at variance with the  random long-range hopping model for which
localization does not occur,
and the average $PR$ never becomes independent of the system size, as it is
shown in Appendix C. 
%Moreover
%the PR of the excited states is
%independent of $N$ for $W>W_{AT}$.  
In the gapped regime  $\gamma > \gamma_{cr}$ where the PR of the excited states 
becomes independent of the system size, the excited
 states display an hybrid nature, with an exponential localized peak
 and a distant independent plateau, in close analogy with the results
 presented in Refs~\cite{alberto,Giulio}.

We note that signatures of localization in presence of homogeneous
long-range hopping
 have already been discussed in Refs~\cite{ossipov,alberto,Giulio}. Here we show
the deep connection  with  the shielding effect: since  the dynamics
of the excited states is shielded from long-range hopping, 
strong signatures of localization can arise.
 
Finally, let us mention that  the ground state  behaves differently: when
$\gamma>\gamma_{cr}$, it becomes delocalized ($PR \propto N$),
%up to a disorder strength which increases with the system size, 
see Fig.~\ref{PRalpha} (lower panel).  
Indeed, in this regime, a simple perturbative argument shows that
the ground state is close to the state $|\psi_N\rangle$,
Eq.~(\ref{SR}), which is extended. 
This implies that in the gapped regime we have a 
 coexistence of extended (ground state)
and localized states  (excited states)  in the same region of parameters.
This might  be used to control transport in 
mesoscopic devices~\cite{kroo}.

\section{Generalization to power law decaying long-range hopping}
In the previous Sections we considered the  case of a distance
independent long-range hopping, Here we present some preliminary
results on the case of a power law decaying long-range hopping. 
We will mainly analyze the localization
properties of the eigenstates and from this analysis we will
conjecture that shielding is a generic properties of homogeneous long-range hopping. 

%\subsection{The energy gap}
The model given in Eq.~(\ref{H}) is modified as follows: 
\begin{multline}
H= D+H_{NN}+V_{LR}=\\
\sum_i E^0_i |i\rangle \langle i| - \Omega \sum_{i} \left(| i \rangle \langle i+1|
+| i+1 \rangle \langle i|\right) \ -\gamma \sum_{i\ne j} \frac{|i\rangle
\langle j|}{|i-j|^{\alpha}}
\label{H2}
\end{multline}
All the terms are the same of Eq.~(\ref{H}), except for 
the last term $V_{LR}$, which now represents 
a long-range hopping term which connects all the sites of the 1-d
chain with an amplitude which decays as a power law with the distance
between the sites $1/r^{\alpha}$. $\alpha$ is the exponent
which determines the range of the interaction: for $\alpha=\infty$ the
interaction involves only  nearest-neighbor sites, while for
$\alpha=0$, the interaction does not decay with the distance and all
the sites are coupled with the same strength, all-to-all interaction.
For $\alpha \le 1$ the hopping is long-range, while for $\alpha >1$
the hopping is short-range, according to the usual definition of the
range of the interaction~\cite{ruffo}. Here, as in Eq.~(\ref{H}), periodic
boundary conditions are assumed both for the nearest neighbor and the
long-range interaction.

Let us first analyze the case $W=0$. 
The most interesting feature of the spectrum of $H_{NN}+V_{LR}$ 
is the presence of an energy gap, $\Delta_0$ between the ground
state and the first excited state in the case of long-range
coupling. While for the case $\alpha=0$ we have shown that
$\Delta_0=N\gamma$, from the exact solutions of the eigenvalue problem, see
Appendix D, we have for any $\alpha$:
\begin{equation}
\Delta_0 \propto \left\{
\begin{array}{cc}
\gamma N^{1-\alpha} &  \quad  {\rm for}  \quad \alpha \le   3  \\
      & \\
      \gamma N^{-2}  &  \quad  {\rm for}  \quad  \alpha> 3  \\

     \end{array}
\right.
\label{Delta0}
\end{equation}

From this expression, we have that the gap increases with the system
size for long-range interaction $\alpha<1$, it is constant for the
critical case $\alpha=1$ and it decreases
with the system size for short-range interaction $\alpha>1$. Note that
the existence of an energy Gap
between the ground state and the first excited state is a peculiar
feature of the long-range non-random hopping interaction. Indeed for
random long-range hopping there is no energy gap.

Clearly the energy gap in the spectrum  will survive even in presence
of disorder if the disorder strength is much smaller than the energy
gap.

%\subsection{Shielding and Localization}

One may wonder whether {\it cooperative shielding} is also
present for generic long-range systems $\alpha <1$.  
For generic long-range hopping, the gap implies that the subspace of
the eigenstates of $V_{LR}$ orthogonal to the fully symmetric
state~Eq.~(\ref{SR}), will become invariant in the large $N$ limit. On the other
side, such subspace
is no longer characterized by only one eigenvalue of $V_{LR}$ as in
the case $\alpha=0$, see analysis in Appendix D. Thus we cannot expect that $H_0$ will alone
determine the dynamics in such subspace. 
We leave to a future work the
detailed analysis of the shielding effect for generic long-range
hopping. Instead here we analyze the localization properties only and
from this analysis we will infer that shielding is present also for
generic long-range hopping. 

\begin{figure}[t!]
\centering
\includegraphics[scale=0.5]{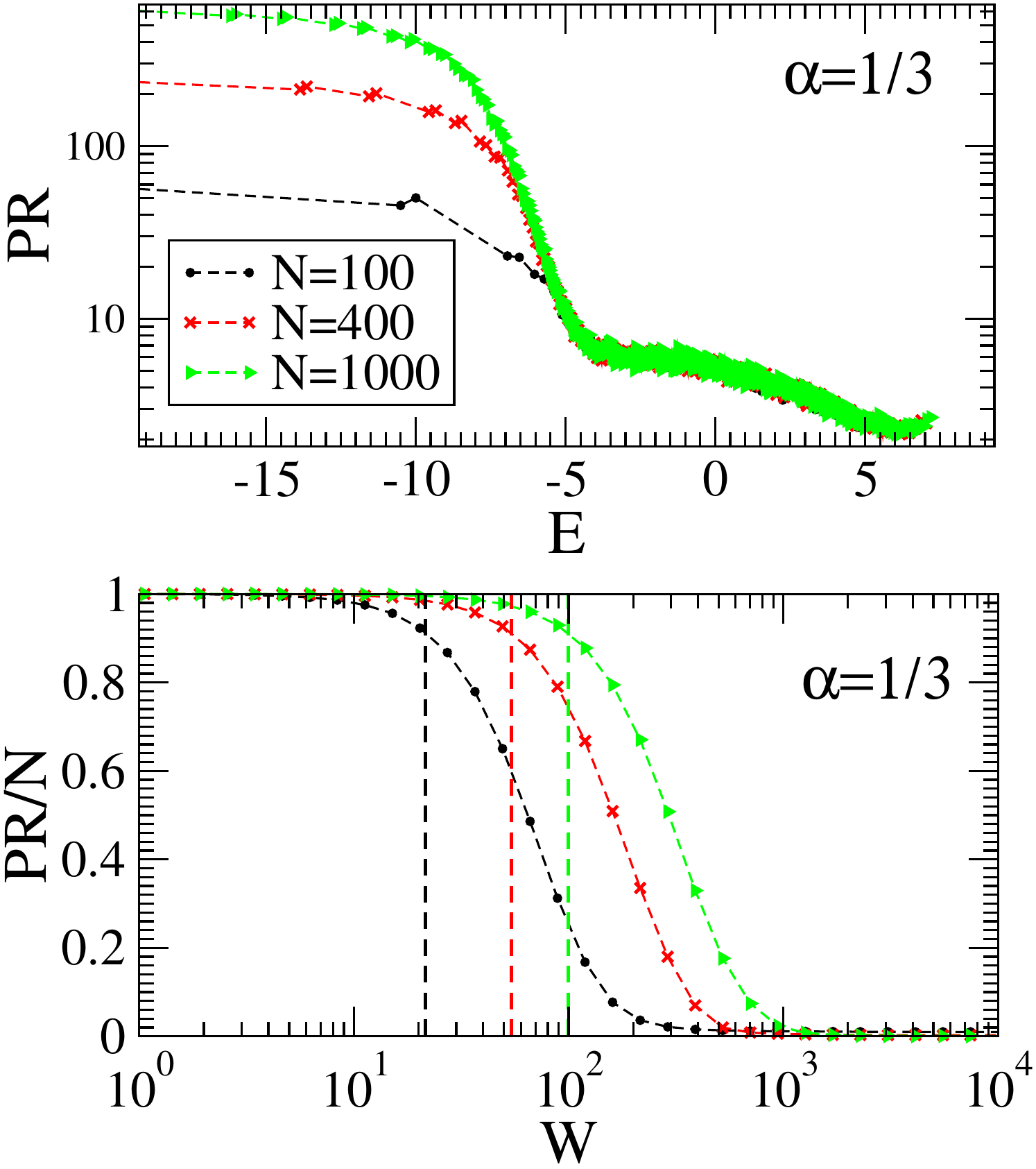}
\caption{Upper panel: Participation ratio is shown {\it vs} energy for the
  eigenstates of the system described by Eq.~(\ref{H2}), for
  different system sizes, see figure. Only the higher energy portion
  of the spectrum is shown. In the lower panel the participation ratio
  of the ground state divided by the number of sites  is shown {\it
    vs} the disorder strength. As a dashed vertical line the estimated
  critical disorder for the onset of localization is shown. The
  critical disorder scales as $\gamma N^{1-\alpha}$.
Parameters used in this figure are: $\Omega=\gamma=1$, $\alpha=1/3$ and $W=10$. 
}
\label{PRalpha2}
\end{figure} 

In  Fig.~\ref{PRalpha2} (upper panel)  the
Participation ratio, see Eq.~(\ref{PR}), averaged over disorder is shown as a function of
energy for the case $\alpha=1/3$. Only the high energy part of the spectrum is shown. The
results  in Fig.~\ref{PRalpha2} (upper panel) show that high
energy states have a participation ratio independent from the system
size, which is a clear sign that features of Anderson localization can
be preserved even in presence of long-range hopping.  
This result is consistent with the conjecture that cooperative shielding
exists also in generic long-range hopping
systems. Indeed in order to have localization we need the long-range
hopping to become effectively short-ranged or to be strongly suppressed
in a given subspace. This
was  the case for $\alpha=0$ where longe-range hopping is strongly
suppressed and it disappears in
the large $N$ limit.

Let us now discuss the localization properties of the ground state. 
In the previous Sections it was also
shown that the ground state for the $\alpha=0$ case  shows cooperative robustness to
disorder, meaning that the disorder needed to localize it
increases with the system size.  Our numerical simulations confirm that
these features are also present for the Hermitian long-range hopping
considered here for generic $\alpha<1$. 
Specifically  disorder must be  larger than the gap
to localize the ground state and we can write for the critical
disorder: $W_{cr} \approx \Delta_0 \propto N^{1-\alpha}$ for
$\alpha<1$. This is  is confirmed in Fig.~\ref{PRalpha} (lower panel) where the
participation ratio of the ground state for the case $\alpha=1/3$ is
shown for different system sizes. 
We have also checked other values of $\alpha<1$ , not shown here.

Thus we have found some general properties of tight binding models with
non-random long-range hopping: a ground state which shows cooperative
robustness to disorder, remaining extended in the large $N$ limit, and
at the same time a part of the excited state subspace which shows strong signature of
localization despite the presence of long-range hopping.

%%%%%%%%%%%%%%%%%%%%%%%%%
%%%%%%%%%%%%%%%%%%%%%%%%%%%%%%%%%%%%%%%%%%%%%%

\section{Conclusions and Perspectives}
The cooperative shielding effect has been analyzed in
a paradigmatic model of single excitation quantum transport, 
described by an Hamiltonian $H=H_0+V_{LR}$, where $H_0$
is the usual Anderson model which contains only nearest-neighbor
hopping and on-site random energies, and  $V_{LR}$ represents a distance
independent homogeneous long-range hopping. 
The homogeneous long-range hopping induces a gap between the ground state and the
excited states of the system, which has the same mathematical 
nature of the superconducting gap
in ultra-small grains. In the regime of large gap,
the dynamics of the
excited states  is described only by $H_0$,
as if long-range would be absent (shielding), up to a time scale which
increases with the system size (cooperativity). 
Thus in the excited state subspace, $H_0$ constitutes an emergent
Hamiltonian, valid up to a finite time scale. 
Such shielding effect  has a strong impact on the transport
properties of the system, allowing for the excited states to show
strong signatures of  localization, 
even in presence of an infinite number of resonances. 
Shielding from long-range  does not depend on the particular form of $H_0$ and it
is determined only by the strength of the long-range hopping term $V_{LR}$. 
We believe that the results presented here  for the infinite 
interaction range, unveil
common cooperative features of long-range hopping systems. 
Indeed our preliminary analysis of generic long-range hopping,
have shown that strong signature of localization arises for any
homogeneous long-range. This result support our conjecture that shielding is
a generic property of long-range hopping.
In perspective it would be interesting to determine the emergent
Hamiltonian which drives the dynamics in the localized part of the
spectrum, for generic long range hopping systems.
Note that the shielding effect in single excitation transport models 
can be currently tested, for instance,  in ion trap experiments.
Moreover the shielding effect can allow for localization to occur in
relevant systems with non random long-range hopping,
such as in cold atomic clouds, where the issue of localization of
light is one of the main open problem.

%%%%%%%%%%%%%%

\section*{Acknowledgments}

We acknowledge useful discussion with R. Bachelard, G. G. Giusteri,
S. Flach, M. Kastner, F. Izrailev, A. M. Mart\'inez-Arg\"uello, E. A. Yuzbashyan, L. Santos.

\newpage
\appendix

\section{Analytical estimate of the shielding time}

\begin{figure}[t!]
\centering
\includegraphics[width=\columnwidth]{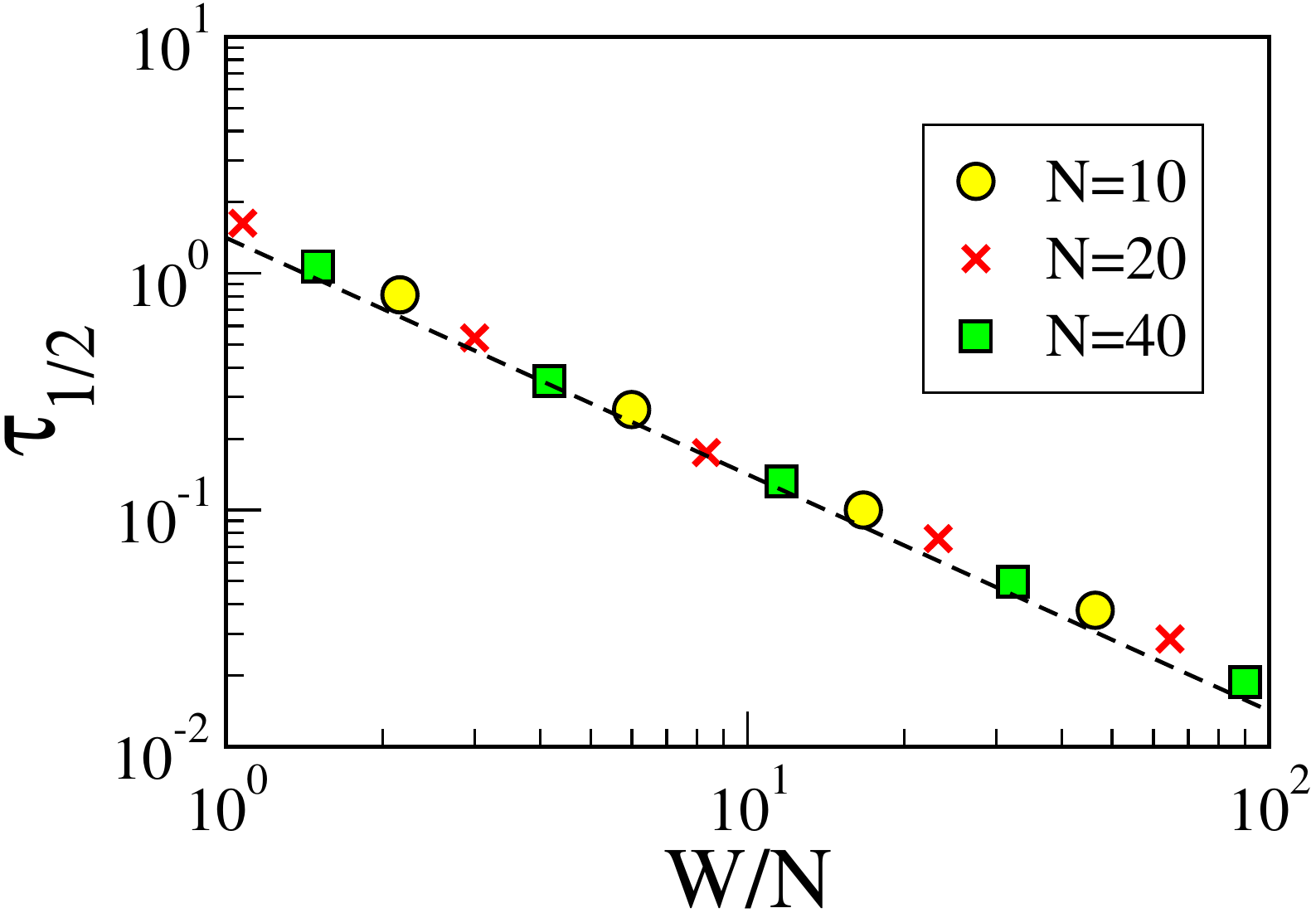}
\caption{Half-time for the Fidelity decay, $\tau_{1/2}$, {\it vs}
the normalized disorder strength $W/N$, for different $N$ values, as indicated 
in the legend. All data are in the regime $\gamma \gg \gamma_{cr}$. 
Symbols are numerical data, line is the analytical prediction
given in Eq.~(\ref{anpr}). Here is $\Omega=1, \gamma=100$.}
\label{tauW}
\end{figure}

Here we estimate the time $\tau_{1/2}$ needed to Fidelity, see Eq.~(\ref{F}), to
decay to $1/2$. We will limit our considerations to the case $W \gg
\Omega$. 
Following Ref.~\cite{Anderson}[17], it is possible to obtain an analytical
expression for  the eigenstates $|E\rangle$ of $H$ in  the regime $\gamma \gg \gamma_{cr}$.
Clearly one
eigenstate is very close to $|\psi_N\rangle$, see Eq.~(\ref{SR})
(due to the presence of a large energy gap).  All the others $N-1$ eigenstates can be written as:
\begin{equation}
\label{appa}
|E \rangle =\frac{1}{\sqrt{C_E }} \sum_{E_0} \frac1{{E -E^0}} |E^0 \rangle\, \hspace {0.1cm}
\end{equation}
with
\begin{equation}
\label{appa1}
\sum_{E_0}\frac{1}{E -E^0}=0.
\end{equation}
and where $C_E $ is a normalization factor. In Eq.~(\ref{appa}), 
$| E_0 \rangle$ and $E_0$ are respectively the eigenstates and eigenvalues of
the Anderson model with Hamiltonian: $H_0=H_{NN}+D$, see Eq.~(\ref{H}). 
From   Eq.~(\ref{appa1})
it is clear that 
the energies $E $ must lie between two
neighbouring energies ($E_0^1,E_0^2$) of $H_0$, so that the main contribution to
the eigenstates $|E \rangle$ comes from the eigenstates of $H_0$
corresponding to such neighbouring energies. 
Thus, approximating 
the initial state as  $|E\rangle \simeq c_1
|E_0^1\rangle + c_2 |E_0^2\rangle$, we have for  the Fidelity,
$$
F_{H_0}= ||c_1|^2+|c_2|^2 e^{i (E_0^1-E_0^2) t} |^2.
$$
Since the mean level distance ($E_0^1-E_0^2$) for
the eigenstates of $H_0$ is $W/N$ (for $W\gg\Omega$), we can estimate
the decay time for the averaged
Fidelity as,
\begin{equation}
\label{anpr}
\tau_{1/2} \propto \frac{N}{W},
\end{equation}
which is confirmed in Fig.~\ref{tauW}. 

\begin{figure}[t!] 
\centering
\includegraphics[width=\columnwidth]{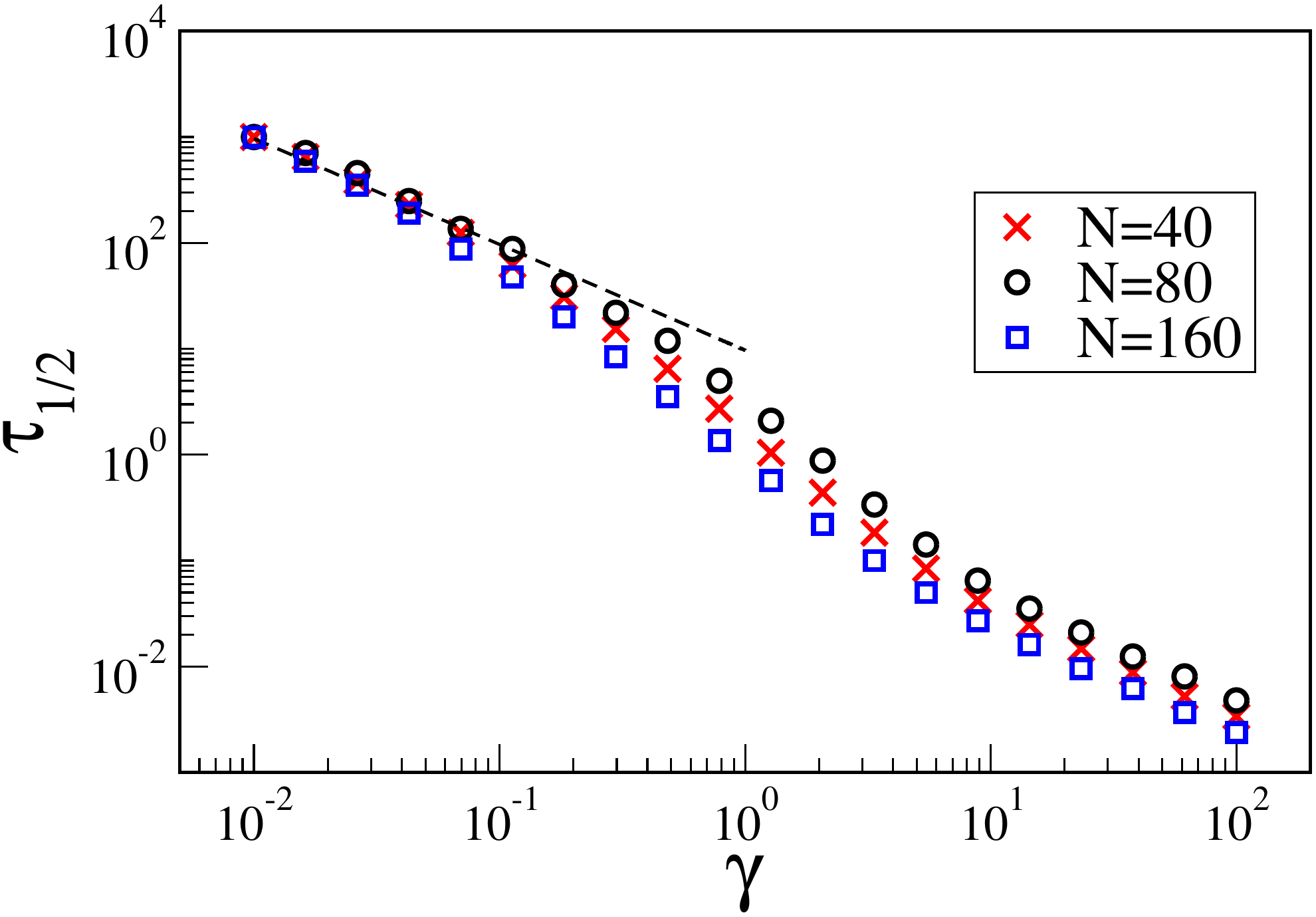}
\caption{
Time needed to the Fidelity to decay to one half, $\tau_{1/2}$ {\it vs} $\gamma$ 
and different $N$ values
for the long-range
random model in  Eq.~(\ref{Hrnd}). 
As initial state we choose a random superposition of all excited states.
Dashed line stands for $C/\gamma$, where $C$ is a fitting constant.
Other data are : $\Omega=1, W=100$ $\alpha=0$. This figure should be
compared with Fig.~\ref{ff} which referred to homogeneous long-range hopping. 
}
\label{ffrnd}
\end{figure}

\begin{figure}[t]
\centering 
\includegraphics[width=\columnwidth]{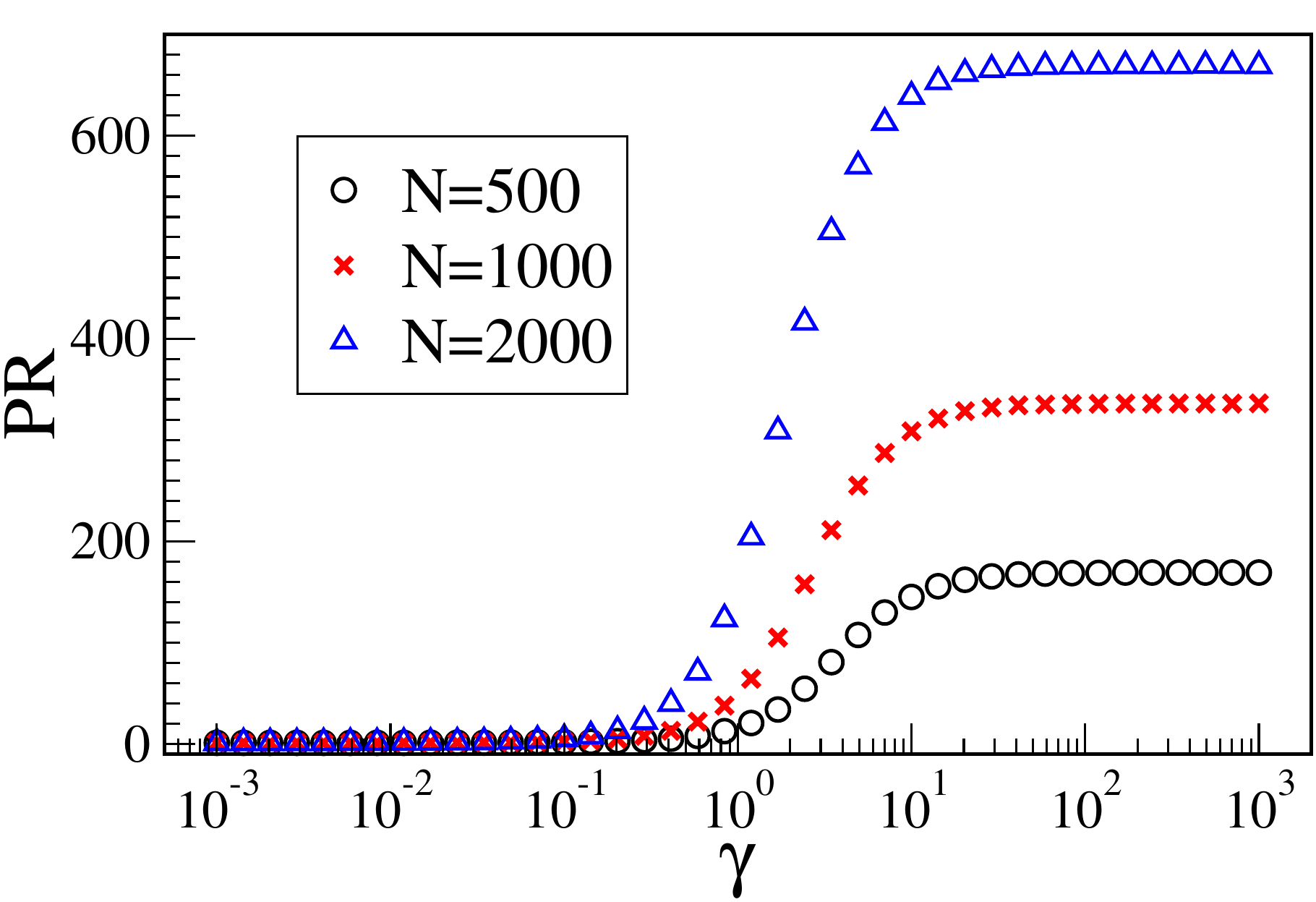}
\caption{Average (over all excited states) participation ratio {\it vs} $\gamma$,
  for
different system size  for the random long-range
hopping described in   Eq.~(\ref{Hrnd}).  
Other parameters  are: $\Omega=1, W=100$. This figure should be
compared with Fig.~\ref{PRalpha} (upper panel). 
}
\label{PRrnd}
\end{figure} 
 
\section{Random hopping model}

In order to show that shielding is due to homogeneous long-range
hopping, here we consider the same model of Eq.~(\ref{H}), but with
random hopping coupling:
\begin{multline}
H_{rnd} = - \Omega \sum_{i} \left( | i \rangle
  \langle i+1| + h.c. \right) + \sum_i E^0_i |i\rangle \langle i| 
\\ -\gamma \sum_{i\ne j} \chi_{i,j} |i\rangle
\langle j| .
\label{Hrnd}
\end{multline}
Here the parameters are all the same of Eq.~(\ref{H}) apart from the
fact that $\chi_{i,j}=\chi_{j,i}$ is a random number in the interval
$(-1/2,+1/2)$.  In this way the long-range coupling is random and it
is easy to show that, in this case,  there is no gap in the spectrum.
This has important consequences on shielding. Indeed for random
coupling the dynamics is never shielded from long-range hopping.
To prove this we have computed the time needed for the  Fidelity, see
Eq.~(\ref{F}), to decay to one half {\it vs} $\gamma$ for different
system sizes $N$. Results are shown in 
Fig.~\ref{ffrnd} where one can observe that $\tau_{1/2}$ always
decays with $\gamma$, in contrast to the case of homogeneous long-range 
hopping where for $\gamma>\gamma_{cr}$ $\tau_{1/2}$ becomes
independent of $\gamma$.  
The absence of shielding in the  random long-range model, also implies
 the absence of localization. This is shown in Fig.~\ref{PRrnd}, where the $PR$ is
 plotted as a function of $\gamma$. As one can see it increases up to
 a saturation point which is proportional to the system size. This is
 at variance with the homogeneous long-range hopping, where the $PR$
 becomes independent of $\gamma$ and $N$ for $\gamma \gg \gamma_{cr}$
 as it is shown in Fig.~\ref{PRalpha}  (upper panel). 

\section{Energy Gap for power-law decaying long-range hopping}
Here we discuss the energy gap $\Delta_0$ between the ground state and the first
excited state in the one dimensional periodic tight binding model
with long-range hopping and no on
site disorder, so that the Hamiltonian given in Eq.~(\ref{H2}) can be written as: 
\begin{multline}
H=- \Omega \sum_i \left( | i \rangle \langle i+1| +| i +1\rangle
   \langle i| \right) 
-\gamma \sum_{i\ne j} \frac{|i\rangle
\langle j|}{r_{i,j}^{\alpha}}
\label{H0}
\end{multline}

For any $\alpha$, $[H_{NN},
V_{LR}]=0$, and the common eigenstates are the same of that of
Eq.~(\ref{H}) for $W=0$. Indeed the  eigenstates $|\psi_q\rangle$ can
be computed exactly~\cite{adame} and their 
 components on the site basis
$|k\rangle$, are independent of $\alpha$ and they can be written as:

\begin{equation}
\braket{k}{\psi_q}=\frac{1}{\sqrt{N}} \exp{\left( i \frac{ 2 \pi k
      q}{N} \right)} \quad
\mbox{with} \quad q=1,...,N
\label{psiq}
\end{equation}

For the eigenvalues of $V_{LR}$, we have~\cite{adame,jul}:
\begin{equation}
E_q^{LR}= \left\{
\begin{array}{lll}
&-2 \gamma \sum_{n=1}^{N/2-1} \frac{\cos{2 \pi q i/N}}{n^{\alpha}}
\quad  & N \quad \mbox{odd} \\
& \\
&-\gamma \frac{(-1)^q}{(N/2)^{\alpha}}-2 \gamma \sum_{n=1}^{N/2-1}
\frac{\cos{2 \pi q i/N}}{n^{\alpha}}     & N \quad
\mbox{even} \\
\end{array}
\right.
\label{EVLR}
\end{equation}
Note that the ground state of  $V_{LR}$ 
for all $\alpha$, corresponds to  $q=N$ and it  is
the fully symmetric and extended state in the site basis, see Eq.~(\ref{SR}).

For the case  $\alpha=0$, $V_{LR}$ has only two
eigenvalues different from zero: one, non degenerate,
corresponding to the ground state, and one, which is $N-1$ degenerate,
corresponding to all the states orthogonal to it. For generic long-range, 
the states orthogonal to the ground state are no longer all
degenerate. 

The eigenvalues of  $H_{NN}$   are: 
\begin{equation}
E_q^{NN}= -2 \Omega \cos{\frac{2 \pi q}{N}} % \text{ with }  q=1,\ldots,N
\label{eigq}
\end{equation}
The common eigenvalues of $H_{NN}+V_{LR}$ are the sum of the 
eigenvalues of the two terms. For the ground state we have $E_N= -2
\Omega + E^{LR}_N$. From these results we can compute the gap (energy
difference between the ground state and the first excited state) $\Delta_0$ for
any $\alpha$ and the result is given in Eq.~(\ref{Delta0}).

\begin{figure}[ht!]
\centering
\includegraphics[scale=0.4]{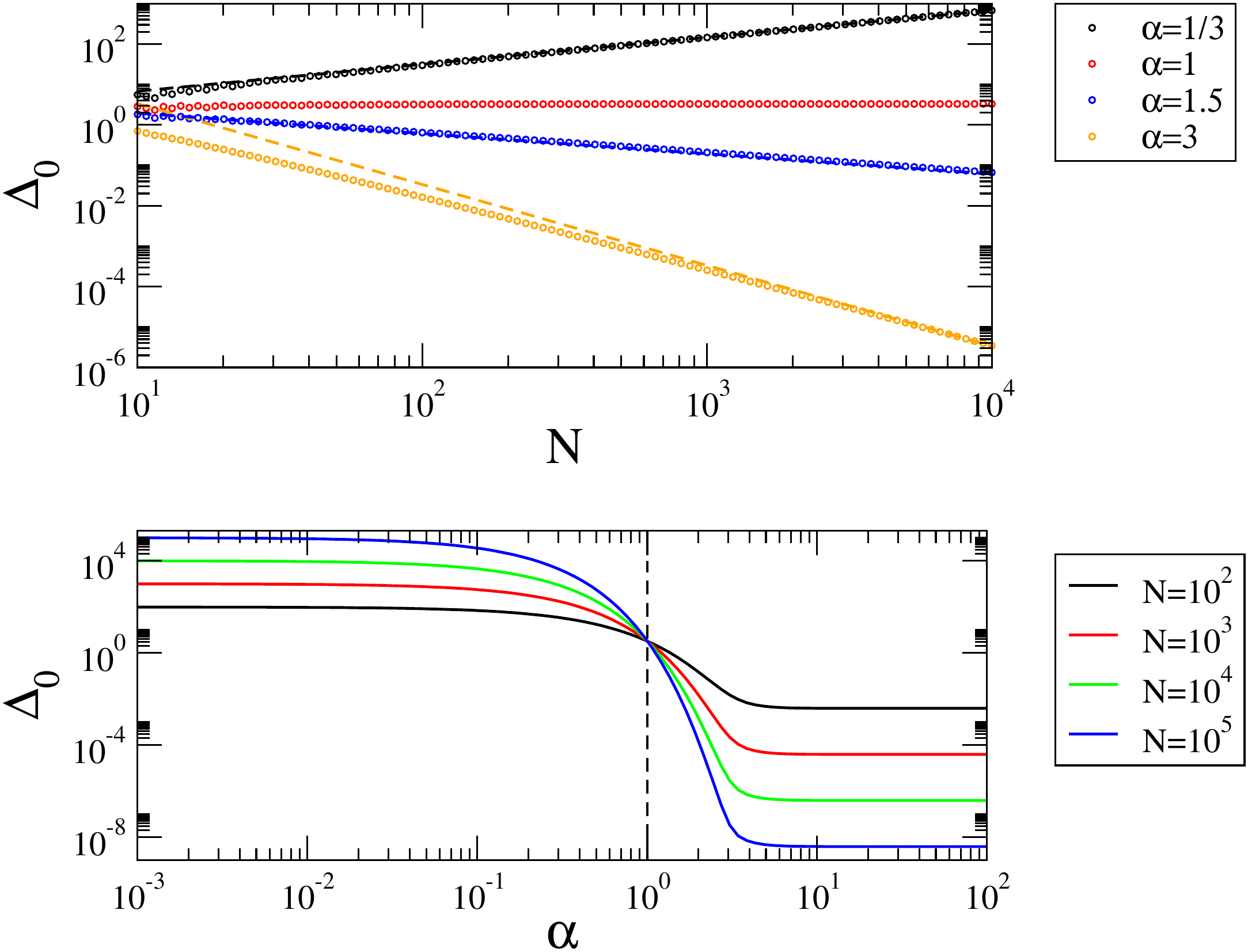}
\caption{Upper panel:  gap between the ground state and the
  first excited state $\Delta_0$  {\it vs} the system size $N$
  for different values of $\alpha$. Dashed lines are the
  asymptotic analytical predictions for $\Delta_0$, see
  Eq.~(\ref{Delta0}). 
Lower panel:
  $\Delta_0$ {\it vs} $\alpha$  for different $N$ values. Data
  refer to the case of no disorder ($W=0$) and
  $\Omega=\gamma=1$, see Eq.~(\ref{H2}). 
}
\label{gapN}
\end{figure}

The analytical prediction about the energy gap is confirmed in Fig.\ref{gapN} .
In Fig.~\ref{gapN} (upper panel) the gap is plotted as a function of the system size
for different values of $\alpha$.  We compare the asymptotic
behaviour given in Eq.~(\ref{Delta0}) with
numerical results, showing that Eq.~(\ref{Delta0}) gives a very good
estimate of the asymptotic behaviour of the gap. 
In Fig.~\ref{gapN} (lower panel) the gap is
plotted as a function of $\alpha$ for different system sizes. 
As one can see for long-range
hopping $\alpha<1$, the gap increases with the system size, it is
constant for the critical case $\alpha=1$, and it decreases with $N$
for $\alpha>1$. 

\vspace{0.5cm}

\end{document}